\documentclass{elsart}

\usepackage[dvips]{graphicx}
\usepackage{amsmath}
\usepackage{amssymb}
\newcommand{\dps}{\displaystyle}
\newcommand{\bfr}{{\mathbf r}}
\newcommand{\bfn}{{\boldsymbol\nabla}}
\newcommand{\po}{\mbox{\it{o}}}
\newcommand{\hf}{\frac{1}{2}}
\newcommand{\pr}[1]{{\sc{\lowercase{#1}}}}
\newcommand{\codeversion}{1.00}
\journal{Computer Physics Communications}
\begin{document}

\begin{frontmatter}

\title{Coordinate-space solution of the Skyrme-Hartree-Fock-Bogolyubov
       equations within spherical symmetry. \\
       The program HFBRAD (v\codeversion)}

\author{K. Bennaceur\corauthref{cor}}
\address{IPN Lyon, CNRS-IN2P3/UCB Lyon 1,
              B\^at. Paul Dirac, \\
              43, Bd. du 11 novembre 1918, 69622 Villeurbanne Cedex,
              France}
\corauth[cor]{Corresponding author.}
\ead{bennaceur@ipnl.in2p3.fr}

\author{J. Dobaczewski}
\address{Institute of Theoretical Physics, Warsaw University     \\
                 ul. Ho\.za 69, PL-00681 Warsaw, Poland                  \\
                 Department of Physics and Astronomy,
                 The University of Tennessee,                            \\
                 Knoxville, Tennessee 37996, USA                         \\
                 Physics Division, Oak Ridge National Laboratory,        \\
                 P.O. Box 2008, Oak Ridge, Tennessee 37831, USA          }
\ead{jacek.dobaczewski@fuw.edu.pl}

\begin{abstract}
We describe the first version (v1.00) of the code \pr{HFBRAD} which
solves the Skyrme-Hartree-Fock or Skyrme-Hartree-Fock-Bogolyubov
equations in the coordinate representation within the sphe\-ri\-cal
symmetry. A realistic representation of the quasiparticle wave
functions on the space lattice allows for performing calculations up
to the particle drip lines. Zero-range density-dependent interactions
are used in the pairing channel. The pairing energy is calculated by
either using a cut-off energy in the quasiparticle spectrum or the
regularization scheme proposed by A.\ Bulgac and Y.\ Yu.
\end{abstract}


\end{frontmatter}

\newpage

{\bf\large PROGRAM SUMMARY}

\textit{Title of the program:} \pr{HFBRAD}
                 (v\codeversion)


\textit{Program obtainable from:}
                      CPC Program Library,
                      Queen's University of Bel\-fast, N. Ireland


\textit{Licensing provisions:} none

\textit{Computers on which the program has been tested:}
                      Pentium-III, Pentium-IV

\textit{Operating systems:} LINUX, Windows

\textit{Programming language used:} FORTRAN-95

\textit{Memory required to execute with typical data:}
30~MBytes

\textit{No. of bits in a word:}
The code is written with a type real corresponding to 32-bit
on any machine. This is achieved using the intrinsic function
{\tt selected\_real\_kind} at the beginning of the code and
asking for at least 12 significant digits.
This can be easily modified by asking for more significant
digits if the architecture of the computer can handle it.

\textit{No.\ of processors used:} 1

\textit{Has the code been vectorised?:} No.

\textit{No.\ of bytes in distributed program, including test data, etc.:}
400~kbytes

\textit{No.\ of lines in distributed program:} 5164
(of which 1635 are comments and separators)

\textit{Nature of physical problem:}
For a self-consistent description of nuclear pair correlations, both
the particle-hole (field) and particle-particle (pairing) channels of
the nuclear mean field must be treated within the common approach,
which is the Hartree-Fock-Bogolyubov theory. By expressing these
fields in spatial coordinates one can obtain the best possible
solutions of the problem; however, without assuming specific
symmetries the numerical task is often too difficult. This is not the
case when the spherical symmetry is assumed, because then the
one-dimensional differential equations can be solved very
efficiently. Although the spherically symmetric solutions are
physically meaningful only for magic and semi-magic nuclei, the
possibility of obtaining them within tens of seconds of the CPU makes them a
valuable element of studying nuclei across the nuclear chart,
including those near or at the drip lines.

\textit{Method of solution:}
The program determines the two-component Hartree-Fock-Bogolyubov
quasiparticle wave functions on the lattice of equidistant points in
the radial coordinate. This is done by solving the eigensystem of two
second-order differential equations by using the Numerov method.
Standard iterative procedure is then used to find self-consistent
solutions for the nuclear product wave functions and densities.

\textit{Restrictions on the complexity of the problem:}
The main restriction is related to the assumed spherical symmetry.

\textit{Typical running time:}
One Hartree-Fock iteration  takes about 0.4~sec for a medium mass
nucleus, convergence is achieved in about 40~sec.

\textit{Unusual features of the program:} none

\vskip 20mm

\goodbreak

{\bf\large LONG WRITE-UP}

\bigskip

\section{Introduction}
\label{sec0}

For several decades, the ground state of nuclei have been studied
within the self-consistent mean-field approximation. Such a
description of the atomic nucleus, has ability to properly account
for the bulk properties of nuclei such as masses, radii or shape. The
Hartree-Fock method provides a good approximation of closed shell
magic nuclei, however, the pairing correlations constitute an
essential ingredient for the description of open shell nuclei. These
effects are usually described by the Hartree-Fock plus BCS (HFBCS) or
Hartree-Fock-Bogolyubov (HFB) methods.

Two main classes of numerical methods have been so far used for
the implementation of the HFBCS or HFB methods. In the first one,
one employs
the expansion of the quasiparticle states on a discrete basis
of orthogonal functions, usually provided by the harmonic
oscillator potential. In this case the non-linear HFB(CS) equations
are formulated in a matrix form and can be solved
by using an iterative procedure (hereafter Bogolyubov iterations).
In the second class, one uses
the direct integration of the equations in the coordinate
representation. This is usually done within the box boundary conditions to
discretize the spectrum of quasiparticle states.
The first approach is more elegant and efficient for
finite-range interactions, however, it has a disadvantage that
the use of harmonic oscillator wave functions may perturb
the correct description of the asymptotics
of the system. On the other hand,
the asymptotic part of the quasiparticle wave functions
does not suffer from this possible weakness when the problem
is solved in the coordinate basis, but here one has to solve
a set of integro-differential equations and as a consequence
this method is usually only applied for the zero-range
Skyrme force, for which the problem reduces to a set
of differential equations.

The question of the asymptotic properties of the atomic nuclei
(neutron skins, halos) is an urgent challenge in theoretical
nuclear structure, and is particularly important due to the
availability of beams of exotic nuclei\footnote{For
example in Europe: ARENAS3, Louvain-la-Neuve; ISOLDE-CERN, Geneva;
GANIL, Caen; and GSI, Darmstadt.}, which allow for relevant experiments.
The HFB problem in coordinate basis, along with the use of the
effective Skyrme force is indeed a powerful tool for
studying ground-state properties of nuclei, and such studies have been and will be
pursued
especially in the regions of nuclear chart where the experiments are being performed.

Two computer codes that solve self-consistent equations
on the basis have recently been published \cite{[Dob04b],[Sto04]}
for non-spherical shapes. We refer the reader to these publications
for a review of approaches and methods that can be and have been
used in such cases. Within the assumed spherical symmetry,
which is the subject of the present study, the first self-consistent
solutions were obtained in the 1970's \cite{[Vau72],[Bei75]}, and
these old codes were later used for decades by different
groups, however, they have never been published.

The first description and implementation of the HFB problem
within the coordinate-space spherical symmetry was presented in Ref.\
\cite{[Dob84]}. The code written during this
work was later updated many times since its early version,
and distributed widely, however, neither this code
was ever the object of an official publication. Indeed, the
absence of accompanying manual makes it hard
to use, and its older versions are sometimes used despite the fact
that the newer ones have been made available.
For coordinate-space spherical symmetry, the code solving the
HFBCS equations for the Skyrme interaction was published in Ref.\ \cite{[Rei91]} and
that for the relativistic mean field Hartree-Bogolyubov method in Ref.\ \cite{[Pos97a]}.
A similar unpublished code also exists to treat pairing correlations for the
finite-range Gogny interaction \cite{[Gra02]}, although the code
that would solve the full HFB problem within these conditions is
not yet available.

The aim of this work is to provide a modern code (named {\tt HFBRAD})
for the spherical Skyrme-HFB problem in the coordinate representation.
The code constitutes a completely rewritten version of that
constructed in Ref.\ \cite{[Dob84]}, but it also features
implementation of the pairing renormalization \cite{[Bul02]}.
Although such code has to be used with caution in open-shell
nuclei, because there the deformation effects not included
here are essential, it nevertheless provides a fast and
easy tool for a quick estimate of nuclear
properties, which can precede much more time
consuming calculations beyond the spherical symmetry.
In this paper we also provide a clear description of the
different Skyrme forces implemented in the code, terms neglected
in the energy functional, and
parameters in the pairing channel.

This paper is organized as follows. In section~\ref{sec1}
we briefly show how the Skyrme-HFB equations are derived
and discuss their main properties. Notably, a significant
part is devoted to the problem of the divergence of the energy
due to the zero range of the force in the pairing channel.
In section~\ref{ch:force} we present several parametrizations of the Skyrme
force implemented in the code, with a particular attention paid to the pairing channel.
In section~\ref{sec4} we give definitions and meanings
of the various quantities and observables which are the result of
calculations. Some aspects of the numerical treatment
of the problem are presented in section~\ref{sec5}, and finally,
sections~\ref{input} and~\ref{output} give a precise description
of the input and output data files.

\section{The Skyrme Hartree-Fock-Bogolyubov equations}
\label{sec1}

The HFB approximation is based on the use of a trial variational
wave function which is assumed to be an independent quasiparticle
state $|\Phi\rangle$. This state, which mixes different
eigenstates of the particle number operator, is a linear combination
of independent particle states representing various possibilities of
occupying pairs of single particle states.
Following the notations and phase convention of~\cite{[Dob84]}
we define the particle and pairing density $\rho$ and
$\tilde\rho$ matrices by
\begin{equation}
\rho(\bfr\sigma q,\bfr'\sigma'q')=\langle\Phi|
a^\dagger_{\bfr'\sigma'q'}a_{\bfr\sigma q}|\Phi\rangle\,,
\end{equation}
\begin{equation}
\tilde\rho(\bfr\sigma q,\bfr'\sigma'q')=-2\sigma'\langle\Phi|
a_{\bfr'-\sigma'q'}a_{\bfr\sigma q}|\Phi\rangle\,,
\end{equation}
where the operators $a^\dagger_{\bfr\sigma q}$ and $a_{\bfr\sigma q}$
create and annihilate a nucleon at the point $\bfr$ having
spin $\sigma=\pm\hf$ and isospin $q=\pm\hf$. The symmetry properties of
$\rho$ and $\tilde\rho$ as well as the relation between
$\tilde\rho$ and the pairing tensor $\kappa$ (defined for example
in~\cite{[RS80]}) are discussed in~\cite{[Dob84]}.

The variation of the energy expectation value
$E=\langle\Phi|\hat{H}|\Phi\rangle$
with respect to $\rho$ and $\tilde\rho$ under the constraints
$N=\langle\Phi|\hat{N}|\Phi\rangle$ and
$Z=\langle\Phi|\hat{Z}|\Phi\rangle$ (for neutrons and protons)
leads to the Hartree-Fock-Bogolyubov equation which reads
in coordinate representation
\begin{multline}
\dps \int d^3{\mathbf r}'\sum_{\sigma'}
\left(
  \begin{matrix}
    h(\mathbf r\sigma,\mathbf r'\sigma')
        & \tilde h(\mathbf r\sigma,\mathbf r'\sigma') \cr
    \tilde h(\mathbf r\sigma,\mathbf r'\sigma')
        & -h(\mathbf r\sigma,\mathbf r'\sigma') \cr
  \end{matrix}
\right)
\left(
  \begin{matrix}
    \varphi_1(E,\mathbf r'\sigma') \cr
    \varphi_2(E,\mathbf r'\sigma') \cr
  \end{matrix}
\right)
= \hskip 2.5cm\hfill\\
\dps \hfill \left(
  \begin{matrix}
    E+\lambda & 0 \cr
    0 & E-\lambda \cr
  \end{matrix}
\right)
\left(
  \begin{matrix}
    \varphi_1(E,\mathbf r\sigma) \cr
    \varphi_2(E,\mathbf r\sigma) \cr
  \end{matrix}
\right)
\end{multline}
where the particle and pairing fields are given by
\begin{equation}
 h(\bfr\sigma,\bfr'\sigma')=
  \frac{\delta E}{\delta\rho(\bfr\sigma,\bfr'\sigma')}\,,\hskip 1.2cm
 \tilde h(\bfr\sigma,\bfr'\sigma')=
  \frac{\delta E}{\delta\tilde\rho(\bfr\sigma,\bfr'\sigma')}\,.
\end{equation}
Once again we refer the reader to the article~\cite{[Dob84]}
for the discussion
concerning the quasiparticle spectrum, its symmetries and the
relations between the components of the HFB spinors and the
densities.

\subsection{Local densities}

In the Skyrme-HFB formalism,
the evaluation of the expectation value of the energy leads
to an expression which is a functional of the local densities,
namely, the particle (normal) and pairing (abnormal) densities
\begin{equation}
\rho(r) = \sum_{i} \varphi_2(E_i,r)^2
\hskip 5mm\mbox{and}\hskip 6mm
\tilde\rho(r) = -\sum_{i} \varphi_1(E_i,r)\varphi_2(E_i,r)\,,
\end{equation}
and their derivatives.
The presence of non local terms in the force leads
to a dependence on the normal and abnormal kinetic densities
\begin{eqnarray}
  \tau(\bfr) &=& \dps
    \sum_{i} \left|\boldsymbol\nabla\varphi_2(E_i,\bfr)\right|^2\,, \hfill\\
  \tilde\tau(\bfr) &=&\dps
    -\sum_{i} \boldsymbol\nabla\varphi_1(E_i,\bfr)\cdot
                \boldsymbol\nabla\varphi_2(E_i,\bfr)\,,
\end{eqnarray}
while the spin-orbit term leads to a dependence on the
spin current tensors ${\mathbb J}_{ij}$ and $\tilde{\mathbb J}_{ij}$.
We do not give here the definitions of these tensors, because for the
spherical symmetry discussed here
they reduce to the corresponding spin
current vectors ${\mathbf J}$ and $\tilde{\mathbf J}$
\begin{eqnarray}
\mathbf J(\bfr)&=&\dps \mathrm{i}\sum_{i}
  \varphi_2(E_i,\bfr)
\bfn\varphi_2(E_i,\bfr)
\langle \sigma'|\hat{\boldsymbol\sigma}|\sigma\rangle, \hfill\\
\tilde{\mathbf J}(\bfr)&=&\dps-\mathrm{i}\sum_{i}
  \varphi_1(E_i,\bfr)
\bfn\varphi_2(E_i,\bfr)
\langle \sigma'|\hat{\boldsymbol\sigma}|\sigma\rangle.
\label{eq:dcs1}
\end{eqnarray}

\subsection{Spherical symmetry}

In the code {\tt HFBRAD} the solutions are restricted
to have a spherical symmetry and do not mix proton
and neutron states. In this situation the
wave functions have the good quantum numbers $(n\ell j m q)$
($n$ is a good quantum number since the spectrum is discretized
inside a spherical box)
and all the solutions inside an $(n\ell j q)$-block are
degenerated. Furthermore, the radial part of the wave functions
can be chosen to be real. Thus we use the ansatz
\begin{equation}
\varphi_i(E,\bfr\sigma)=\frac{u_i(n\ell j,r)}{r}
{\mathrm Y}^{(\ell)}_{m_\ell}(\hat r)
\langle \ell m_\ell\,{\textstyle\hf}\sigma|jm\rangle\,,\hskip 1cm i=1,\,2\,,
\label{eq:spherwf}
\end{equation}
for the wave functions. The local densities can be written
using the radial functions (omitting the isospin quantum number)
\begin{eqnarray}
\rho(r)&=&\dps
  \frac{1}{4\pi r^2}\sum_{n\ell j}(2j+1)u_2^2(n\ell j,r)\,,\hfill\\
\tilde\rho(r)&=&\dps
  -\frac{1}{4\pi r^2}\sum_{n\ell j}(2j+1)u_1(n\ell j,r)
  u_2(n\ell j,r)\,.\hfill
\end{eqnarray}
For the kinetic densities we have
\begin{eqnarray}
\tau(r)&=&\dps
  \sum_{n\ell j}\frac{2j+1}{4\pi r^2}
  \left[
    \left(u_2'(n\ell j,r)-\frac{u_2(n\ell j,r)}{r}\right)^2
    +\frac{\ell(\ell+1)}{r^2}u_2^2(n\ell j,r)
  \right]\,,
  \hfill\\
\tilde\tau(r)&=&\dps
  -\sum_{n\ell j}\frac{2j+1}{4\pi r^2}
  \left[
    \left(u_1'(n\ell j,r)-\frac{u_1(n\ell j,r)}{r}\right)
    \left(u_2'(n\ell j,r)-\frac{u_2(n\ell j,r)}{r}\right) \right.
          \nonumber\hfill\\
    &&\hfill\dps+\left.\frac{\ell(\ell+1)}{r^2}u_1(n\ell j,r)u_2(n\ell j,r)
  \right]\,.
\end{eqnarray}
Finally the spin current vector densities have only one non vanishing
component given by
\begin{eqnarray}
J(r) &=&\dps \frac{1}{4\pi r^3}\sum_{n\ell j}(2j+1)
  \left[j(j+1)-\ell(\ell+1)-\frac{3}{4}\right]u_2^2(n\ell j,r)\,,\hfill\\
\tilde J(r) &=&\dps-\frac{1}{4\pi r^3}\sum_{n\ell j}(2j+1)
  \left[j(j+1)-\ell(\ell+1)-\frac{3}{4}\right]u_1(n\ell j,r)u_2(n\ell j,r)\,.
\end{eqnarray}

\subsection{The Hartree-Fock-Bogolyubov energy}

In the Skyrme-HFB approximation, the total energy $\mathcal{E}$ of a
nucleus is given as a sum of kinetic, Skyrme, pairing and Coulomb
terms:
\begin{eqnarray}
E&=& K+E_{\mathrm{Skyrme}}
  +E_{\mathrm{pair}}+ E_{\mathrm{Coul}}\hfill\nonumber\\
 &=&\!\!\int\!d^3\bfr\,\left[\mathcal{K}(\bfr)
        +\mathcal{E}_{\mathrm{Skyrme}}(\bfr)
  +\mathcal{E}_{\mathrm{pair}}(\bfr)+\mathcal{E}_{\mathrm{Coul}}(\bfr)\right].
\label{eq:etot}
\end{eqnarray}
The derivation of the energy is explained in detail in several
articles (see Refs.\ \cite{[Vau72],[Eng75]} for the HF energy
and~\cite{[Dob84]} for the HFB case). The expression of the Skyrme
force is given in section~\ref{ch:force}.

In equation~(\ref{eq:etot}),
the kinetic energy of both neutrons and protons is given by
isoscalar kinetic density. The neutron and proton
masses being approximated by their average value, and one has
\begin{equation}
\mathcal{K}=\frac{\hbar^2}{2m}\tau\left(1-\frac{1}{A}\right)\,,
\end{equation}
where the factor in parentheses takes into account the
direct part of the center-of-mass correction \cite{[Ben00]}.
Since we only consider even-even nuclei, the Skyrme part of the
energy functional is time even, it can be written as a sum of
isoscalar ($T=0$) and isovector ($T=1$) parts (see {\it e.g.}
Ref.\ \cite{[Dob95]})
or as a sum on isoscalar, neutrons and protons densities,
\begin{eqnarray}
\!\!\!\!\mathcal{E}_{\mathrm{Skyrme}}&=&\dps
  \hf t_0\left[\left(1+\frac{x_0}{2}\right)\rho^2-\left(x_0+\hf\right)
  \sum_q\rho_q^2\right]\hfill\nonumber \\
  &+&\dps
  \frac{t_1}{4}\left\{\left(1+\frac{x_1}{2}\right)\left[\rho\tau+\frac{3}{4}
    \left(\nabla\rho\right)^2\right]-\left(x_1+\hf\right)
    \sum_q \left[\rho_q\tau_q+\frac{3}{4}\left(\nabla\rho_q\right)^2\right]
  \right\}\hfill\nonumber \\
  &+&\dps
  \frac{t_2}{4}\left\{\left(1+\frac{x_2}{2}\right)\left[\rho\tau-\frac{1}{4}
    \left(\nabla\rho\right)^2\right]+\left(x_2+\hf\right)
    \sum_q \left[\rho_q\tau_q-\frac{1}{4}\left(\nabla\rho_q\right)^2\right]
  \right\}\hfill\nonumber \\
  &-&\dps
  \frac{1}{16}\left(t_1x_1+t_2x_2\right)J^2
  +\frac{1}{16}\left(t_1-t_2\right)\sum_q J_q^2 \hfill\nonumber \\
  &+&\dps
  \frac{1}{12}t_3\rho^\gamma
    \left[\left(1+\frac{x_3}{2}\right)\rho^2-\left(x_3+\hf\right)
    \sum_q\rho_q^2
    \right]\hfill\nonumber \\
  &+&\dps
    \hf W_0\left(J\nabla\rho+\sum_q J_q\nabla\rho_q\right)  \label{skyrme:e}
\end{eqnarray}
Index $q$ stands for neutrons and protons while the absence
of index indicates the total (isoscalar) density.
Using the same notations for the densities, the pairing energy density reads
\begin{eqnarray}
\mathcal{E}_{\mathrm{pair}}=\dps
  \sum_q \left\{ \frac{t_0'}{4}(1-x_0')\tilde\rho^2_q
  +\frac{t_1'}{4}(1-x_1')\left[
   \tilde\rho_q\tilde\tau_q+\frac{1}{4}\left(\nabla\tilde\rho_q\right)^2
   \right]\right.\ \ \ \ &\nonumber\\
 \dps \hfill\left.+\left[\frac{t_2'}{8}(1+x_2')+
  \frac{1}{4} W_0'\right]\tilde J_q^2
  +\frac{t_3'}{24}(1-x_3')\rho^{\gamma'}\tilde\rho_q^2\right\}&
\end{eqnarray}
In this last expression we have added a prime to the parameters
of the interaction. Although the derivation of the general
Skyrme energy density functional is based on a unique force,
its {\it effective} nature justifies the use of different
sets of parameters in the particle-hole and particle-particle
channels. The prime indices anticipates this possibility
which is discussed in section~\ref{ch:force}.

The energy density defined in (\ref{eq:etot}) involves
a Coulomb term for protons. This term contains a direct
part which can be expressed using the charge density and
leads to a local field after variation of the energy and an
exchange part which would lead to a non local potential. Both
parts deserve a special discussion.
The direct part of the
Coulomb energy depends on the charge density $\rho_{ch}(\bfr)$ and reads
\begin{equation}
\mathcal{E}_{\mathrm{coul}}^{\mathrm{dir}}
  =\frac{e^2}{2}\int\!\!\!\int d^3{\bfr}\,d^3{\bfr'}\,
\frac{\rho_{ch}(\bfr)\rho_{ch}(\bfr')}{|\bfr-\bfr'|}\,.
\label{eq:ecdir}
\end{equation}
We use the point proton density to simplify this expression,
{\it i.e.} the charge density is replaced by the proton density
$\rho_p(\bfr)$. Nevertheless, the proton form factor is taken
into account when we calculate the charge radius of the nucleus
(see eq.~(\ref{eq:rcharge})).
The exchange part of the Coulomb energy leads to a non local term
and is treated with the Slater approximation. This approximation
means that we keep only the first term of the density matrix
expansion in the local density approximation~\cite{[Neg72]}
\begin{equation}
\mathcal{E}_{\mathrm{coul}}^{\mathrm{ex}}=
  -\frac{3}{4}e^2\left(\frac{3}{\pi}\right)^{\frac{1}{3}}
\int\! d^3\bfr\, \rho_p^{4/3}(\bfr)\,.
\label{eq:ecex}
\end{equation}
The error introduced by this approximation
has been estimated by studying the next order
term~\cite{[Tit74]} or more recently by comparison with the
exact treatment of the Coulomb energy~\cite{[Ska01]}.
The Slater approximation seems to have little consequence on the nuclei
bulk properties although some significant effects can be expected
on the position of the proton drip-line and the Coulomb displacement
energy of single particle levels.
Finally, the Coulomb contributions to the pairing energy and fields
are not included in the {\tt HFBRAD} code.

\subsection{The Hartree-Fock-Bogolyubov mean fields}

Since the Bogolyubov transformation does not preserve the particle
number, we introduce two Lagrange multipliers $\lambda_N$ and
$\lambda_Z$ to conserve the average neutron and proton number.
The HFB equations are then obtained by writing the stationary
condition
$\delta \left[\mathcal{E}-\langle\lambda_N N +\lambda_Z Z\rangle\right]=0$.
The dependence of $\mathcal{E}$ on the kinetic densities leads
the effective mass
\begin{eqnarray}
M_q&=&\dps\frac{\hbar^2}{2m^*_q}\hfill\\
 &=&\dps\frac{\hbar^2}{2m}+\frac{t_1}{4}
 \left[\left(1+\frac{x_1}{2}\right)\rho
  -\left(x_1+\hf\right)\rho_q\right]\hfill\nonumber\\
&&\hskip 1.4cm +\frac{t_2}{4}\left[\left(1+\frac{x_2}{2}\right)\rho
  +\left(x_1+\hf\right)\rho_q\right]
\end{eqnarray}
and to the abnormal effective mass
\begin{equation}
\tilde M_q=\frac{t_1'}{4}(1-x_1')\tilde\rho_q\,.
\end{equation}
The particle-hole (Hartree-Fock) fields is given by
\begin{eqnarray}
U_q&=&\dps
  t_0\left[\left(1+\frac{x_0}{2}\right)\rho
       -\left(x_0+\hf\right)\rho_q\right] \hfill\nonumber \\
&+&\dps
 \frac{t_1}{4}\left[\left(1+\frac{x_1}{2}\right)
  \left(\tau-\frac{3}{2}\Delta\rho\right)
 -\left(x_1+\hf\right)\left(\tau_q-\frac{3}{2}\Delta\rho_q\right)
  \right] \hfill\nonumber \\
&+&\dps
 \frac{t_2}{4}\left[\left(1+\frac{x_2}{2}\right)
  \left(\tau+\hf\Delta\rho\right)
 +\left(x_2+\hf\right)\left(\tau_q+\hf\Delta\rho_q\right)
  \right] \hfill\nonumber \\
&+&\dps
  \frac{t_3}{12}\left[\left(1+\frac{x_3}{2}\right)(2+\gamma)
    \rho^{\gamma+1}
  -\left(x_3+\hf\right)\left(\gamma\rho^{\gamma-1}\sum_{q'}\rho_{q'}^2
  +2\rho^\gamma\rho_q\right)\right] \hfill\nonumber \\
&+&\dps
  \frac{t_3'}{24}(1-x_3')\gamma'\rho^{\gamma'-1}
  \sum_{q'}\tilde\rho_{q'}^2 \hfill\nonumber \\
&-&\dps
  \frac{W_0}{2}\left(\nabla J+\nabla J_q\right)
\end{eqnarray}
In the case of protons,
varying the expressions~(\ref{eq:ecdir})
and~(\ref{eq:ecex}) leads to the following expression for the Coulomb
field
\begin{equation}
V_c(\bfr)=\frac{e^2}{2}\int\, d^3\bfr'\,\frac{\rho_p(\bfr')}{|\bfr-\bfr'|}
-e^2\left(\frac{3}{\pi}\right)^{\frac{1}{3}}\rho_p^{1/3}(\bfr)\,.
\end{equation}
The particle-particle (pairing) field is
\begin{equation}
\tilde U_q=
\frac{t_0'}{2}(1-x_0')\tilde\rho_q
+\frac{t_1'}{4}(1-x_1')\left[\tilde\tau_q-\hf\Delta\tilde\rho_q\right]
+\frac{t_3'}{12}(1-x_3')\rho^{\gamma'}\tilde\rho_q\,.
\end{equation}
Finally, the spin-orbit fields ($\propto\boldsymbol\ell\cdot\mathbf s$)
have the following form factors:
\begin{equation}
B_q=-\frac{1}{8}\left(t_1x_1+t_2x_2\right)J+\frac{1}{8}
(t_1-t_2)J_q+W_0\nabla(\rho+\rho_q)\,,
\label{mf:so}
\end{equation}
\begin{equation}
\tilde B_q=\left[\frac{t_2'}{2}(1+x_2')+W_0'\right]\tilde J_q\,.
\end{equation}

\subsection{The Hartree-Fock-Bogolyubov equations}
\label{sec:hfbeq}

Writing the fields in matrix form
\begin{equation}
\mathcal{M}=\left(\begin{matrix}
M& \tilde M \\
\tilde M  & - M \\
\end{matrix}\right)\,,
\hskip 1.3cm
\mathcal{U}=\left(\begin{matrix}
U-\lambda& \tilde U \\
\tilde U  & - U+\lambda \\
\end{matrix}\right)\,,
\end{equation}
\begin{equation}
\mathcal{U}_{\mathrm{so}}=\left(\begin{matrix}
B & \tilde B \\
\tilde B  & - B \\
\end{matrix}\right)
\frac{j(j+1)-\ell(\ell+1)-\frac{3}{4}}{2r}
\,.
\end{equation}
The HFB equations read
\begin{equation}
\left[-\frac{d}{dr}\mathcal{M}\frac{d}{dr}+\mathcal{U}+\mathcal{M}
\frac{\ell(\ell+1)}{r^2}+\frac{\mathcal{M}'}{r}
+\mathcal{U}_{\mathrm{so}}\right]
\left(\begin{matrix}
u_1\\u_2\\
\end{matrix}\right)
=E
\left(\begin{matrix}
u_1\\u_2\\
\end{matrix}\right)\,.
\label{eq:sysfin}
\end{equation}
An $r$-dependant mixing of components and scaling described
in Ref.\ \cite{[Dob84]} allows us to write this
equation as an equation with no differential operator in
the coupling terms and no first order derivative
\begin{equation}
\label{HFBdiff}
\begin{array}{rcl}
-M^*\frac{d^2}{dr^2}f_1+Vf_1+Wf_2 &=& Ef_1 ,\\
M^*\frac{d^2}{dr^2}f_2-Vf_2+Wf_1 &=& Ef_2  .
\end{array}
\end{equation}
This last form with no first order derivative of the functions
is particularly suitable for the numerical integration by the
Numerov algorithm briefly discussed in section~\ref{sec:numerov}.

\subsection{Asymptotic properties of the HFB wave functions}

The asymptotic properties of the two components of the HFB
quasiparticle wave functions, and their dependence on $\lambda$ and
$E$, were discussed in Refs.~\cite{[Bul80],[Dob84],[Dob96]}. Here we
complement this discussion by further elements, which pertain mainly
to weakly bound systems where the Fermi energy $\lambda$ is very
small.

Assuming that $M^*$=$\frac{\hbar^2}{2m}$, which anyhow is always
fulfilled in the asymptotic region, and neglecting for simplicity
this trivial mass factor altogether, Eqs.~(\ref{HFBdiff}) have for
large $r$ the following form
\begin{equation}
\begin{matrix}
f_1'' &+& (E+\lambda)f_1 &=& W f_2 , \\
f_2'' &-& (E-\lambda)f_2 &=& -Wf_1 . \\
\end{matrix}
\label{HFBdiff1}
\end{equation}
Here we have also neglected all mean-field potentials, assuming that
the distance is large enough that they are much smaller than
the two characteristic constants $k^2=E+\lambda$ and $\kappa^2=E-\lambda$.
Moreover, we consider the case of small Fermi energy such that
there are no discrete HFB states, i.e., $E$$\geq$$-\lambda$.
Within such conditions, the only remaining question is whether in the
asymptotic region we can neglect the coupling potential $W$.

In order to discuss this question, we note that the coupling terms
can be considered as inhomogeneities of the linear
equations~(\ref{HFBdiff1}), and therefore, asymptotic solutions
have the form
\begin{equation}
\begin{matrix}
f_1(r) &\propto& \sin(kr+\delta) + F_1(r) , \\
f_2(r) &\propto& e^{-\kappa r} +   F_2(r)   . \\
\end{matrix}
\label{eq:asympt}
\end{equation}
We can now iterate these solutions starting with $F_1(r)$=$F_2(r)$=0,
which corresponds to neglecting the coupling terms in the zero order.
Then, in the first order we see that in the equation for $f_1$ the
coupling term can still be neglected as compared to the large
term $(E+\lambda)f_1$. Therefore, to all orders we have $F_1(r)$=0,
and the asymptotic solution $f_1(r)\propto\sin(kr+\delta)$ defines
the inhomogeneity $Wf_1$ of the equation for $f_2$.

We arrive here at the conclusion that the asymptotic form of $f_2$
may involve two terms and a more detailed analysis is needed before
concluding which one dominates. To this end, we note that the coupling
potential $W$ depends on the sum of products of lower and upper
components, and therefore has a general form of
\begin{equation}
W(r) \propto W_{\mathrm{osc}}(r) e^{-\mu r}/r^2 ,
\end{equation}
where $W_{\mathrm{osc}}(r)$ is an oscillating function of order 1,
$\mu$ is the decay constant, which may characterize another
quasiparticle state than the one with energy $E$, and the factor $r^2$
comes from the volume element, cf.~Eq.~(\ref{eq:spherwf}). We see
here that the term $F_2(r)$ in the asymptotic form of $f_2(r)$
vanishes as $e^{-\mu r}$, but it is difficult to say
which one of the two terms dominates asymptotically. We only note that
for small quasiparticle energies the decay constant $\kappa$ is
also small and hence the first term is then more likely to dominate.

\begin{figure}
\begin{center}
$\begin{matrix}
\includegraphics*[scale=0.45,clip,bb=62 53 490 340]{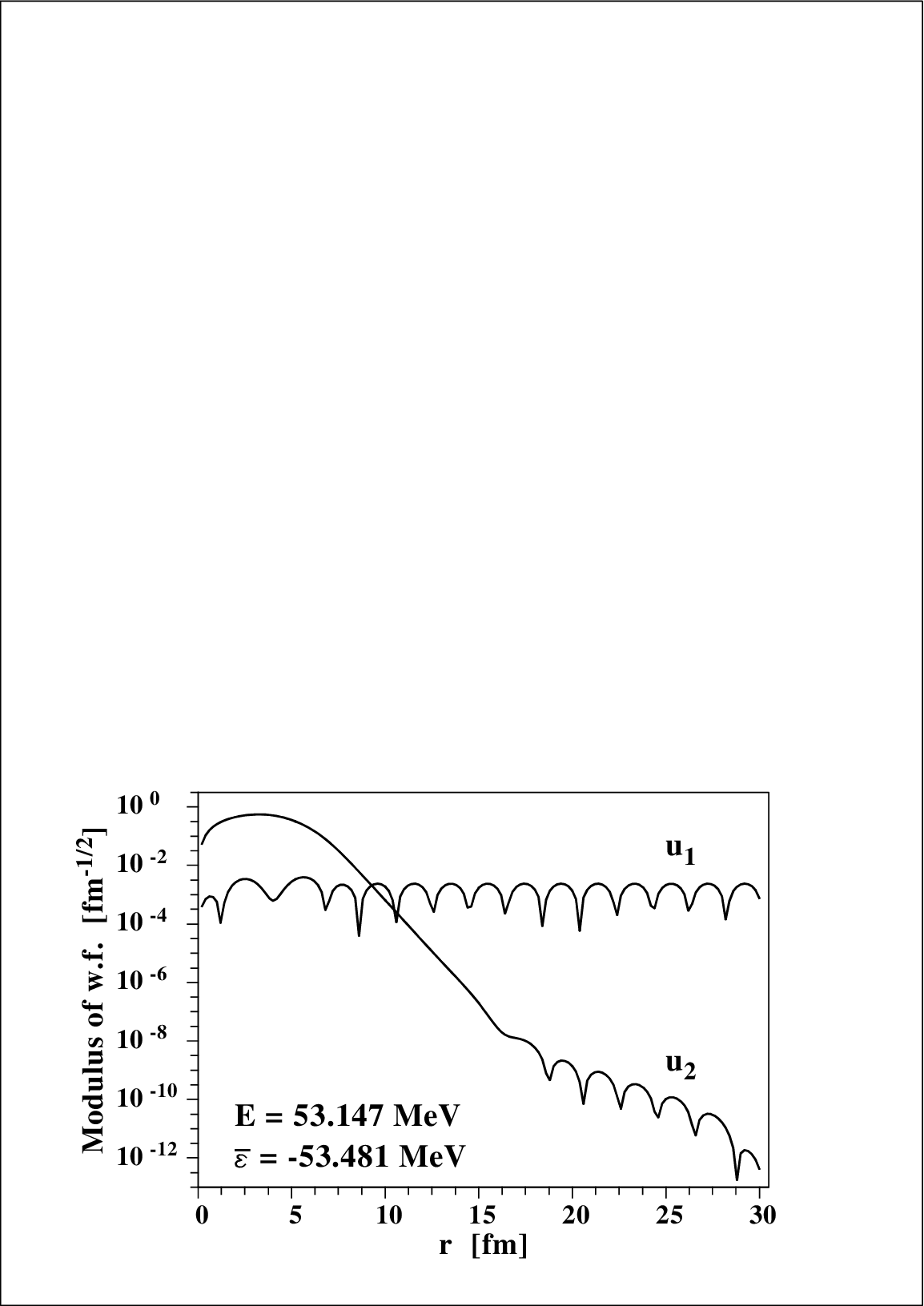} &
\includegraphics*[scale=0.45,clip,bb=62 53 490 340]{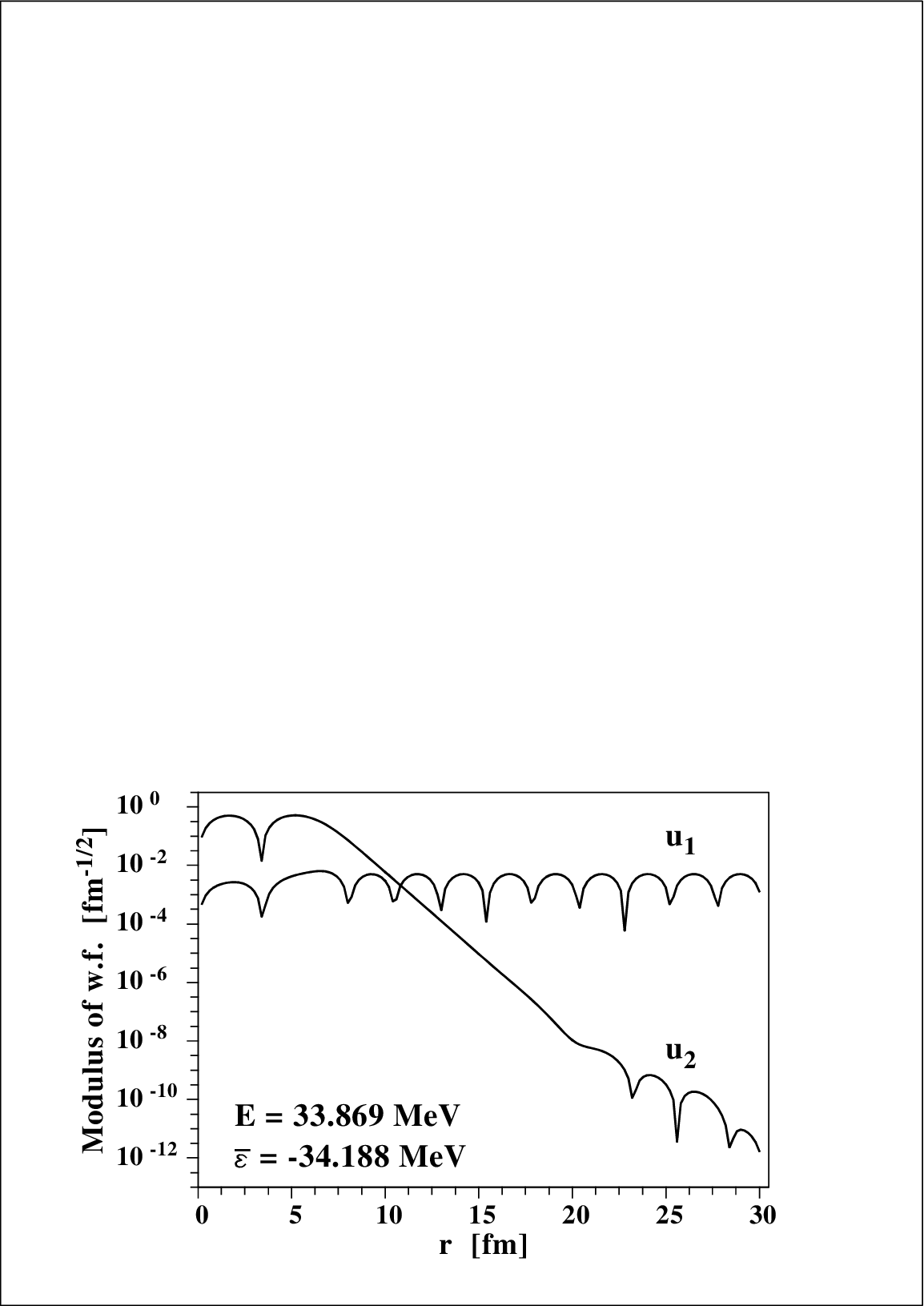} \\
\includegraphics*[scale=0.45,clip,bb=62 53 490 340]{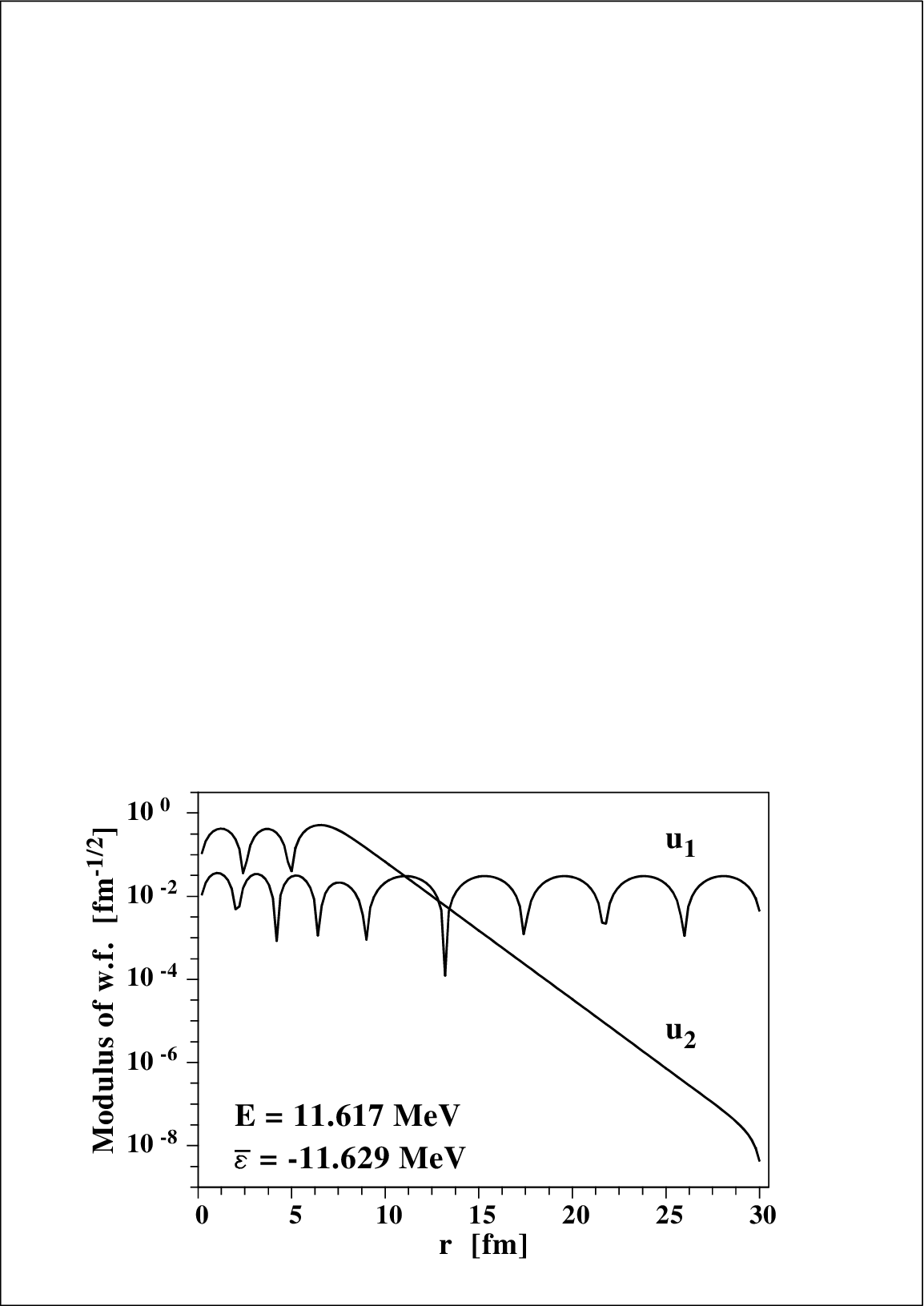} &
\includegraphics*[scale=0.45,clip,bb=62 53 490 340]{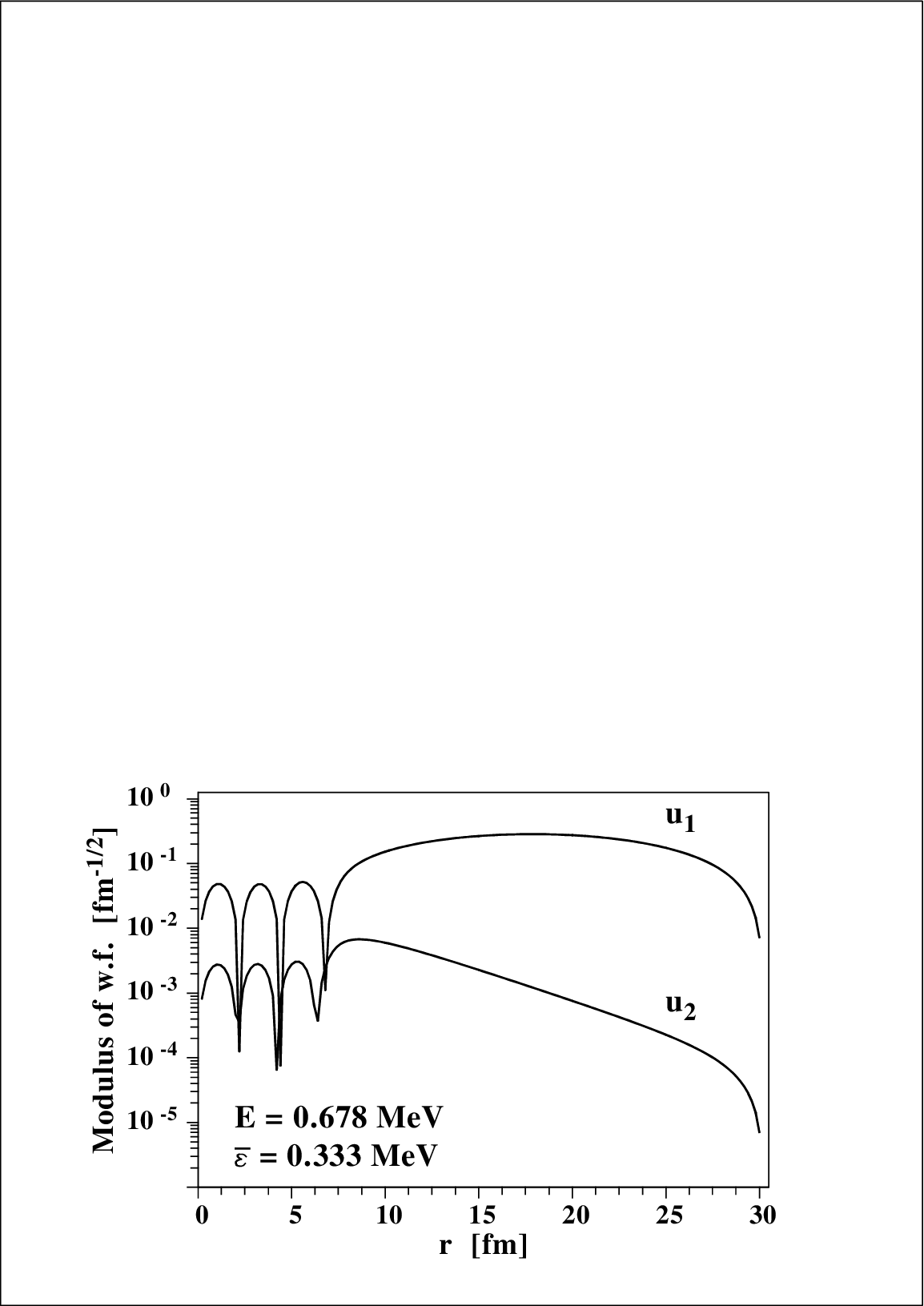} \\
\end{matrix}$
\caption{HFB quasiparticle wave functions $u_1(r)$ and $u_2(r)$
of the four lowest states in $^{174}$Sn.
\label{figc}}
\end{center}
\end{figure}

In order to illustrate the above discussion, we have performed the
HFB calculations in $^{174}$Sn, where $\lambda_n$=$-$0.345\,MeV and
the lowest $\ell$=0 quasiparticle state of $E$=0.429\,MeV  (for
$R_{\mathrm{box}}$=30\,fm) leads to a very diffused coupling
potential with small decay constant $\mu$. Quasiparticle wave
functions corresponding to the four lowest $\ell$=0 quasiparticle
states are shown in Fig.~\ref{figc}. One can see that the asymptotic
forms of the second components  $u_2(r)$ of the two lowest
quasiparticle states are not affected by the second term $F_2(r)$, at
least up to 30\,fm. Only the third and fourth states switch at large
distances to the oscillating asymptotic forms with a smaller decay
constant. This happens at rather large distances where the densities
are anyhow very small. Hence the change in the asymptotic properties
does not affect any important nuclear observables.

In all cases that we have studied, the practical importance of the
second term $F_2(r)$ is negligible. However, its presence precludes
simple analytic continuation of the wave functions in the asymptotic
region. We would like to stress that the asymptotic forms discussed
in this section are numerically stable, and that they are unrelated
to the numerical instabilities discussed in Sec.~\ref{sec:instab} below.

\subsection{Pairing correlations and divergence of the energy}
\label{sec2.6}

The use of a zero range interaction in the particle-particle channel
leads to the energy divergence problem~\cite{[Bul02],[Bul01c]} related to the
fact that the short-range behaviour of the abnormal density is
\begin{equation}
\tilde\rho(r)=\lim_{r'\rightarrow r}\tilde\rho(r,r')
\propto\frac{1}{|r-r'|} .
\end{equation}
The other abnormal densities diverge as well, but along that
the pairing correlations also
induce the divergence of the kinetic density $\tau(r)$.
One simple way to overcome the divergence of the energy is
to introduce a cut-off energy $E_{\mathrm{cut}}$ in the summations of all the densities
\begin{equation}
\sideset{}{_i}\sum_{E_i>0}
\ \rightarrow\ \sideset{}{_i}\sum_{0<E_i<E_{\mathrm{cut}}}\,.
\end{equation}
This cut-off somehow simulates the finite range of the interaction in
the pairing channel. It therefore constitutes
an additional parameter of the interaction
and must be adjusted along with the other quantities which define
the pairing interaction, see e.g.\ Ref.\ \cite{[Dob96]}.

A more elegant way to prevent the divergence was recently
proposed in Refs.\ \cite{[Bul02],[Bul01c]}. It is based on the
subtractions of the divergent part of the abnormal density.
In its present implementation in the code {\tt HFBRAD},
this method can only be applied if the pairing force
does not contains non-local terms ({\it i.e.} gradient
terms). In that case, the pairing energy depends only
on $\tilde\rho$ (and possibly on $\rho$ but not on $\tilde\tau$
or $\tilde J$), i.e.,
\begin{equation}
\mathcal{E}_{\mathrm{pair}}=\sum_q
\left[\frac{t_0'}{4}(1-x_0')+\frac{t_3'}{24}(1-x_3')\rho^{\gamma'}\right]
\tilde\rho_q^2=\sum_q g[\rho]\tilde\rho_q^2 ,
\end{equation}
and the pairing potential depends linearly on $\tilde\rho$,
\begin{equation}
\tilde U_q=2 g[\rho]\tilde\rho_q\,.
\end{equation}
The regularized pairing field can then be written as
\begin{equation}
\tilde U^{\mathrm{reg}}_q=2 g[\rho]\tilde\rho_q^{\mathrm{reg}}
\equiv2 g_{\mathrm{eff}}[\rho]\tilde\rho_q\,.
\end{equation}
Calculation of the various densities is performed by summing up
the quasiparticle contributions up to a maximum energy $E_{\mathrm{max}}$, but because
of the regularization, this maximum energy has not the meaning
of a cut-off energy $E_{\mathrm{cut}}$.

It can be shown \cite{[Bul01c]}
that when we evaluate the total energy of the system,
the pairing energy must be defined as
\begin{equation}
\mathcal{E}_{\mathrm{pair}}=\sum_q g_{\mathrm{eff}}[\rho]\tilde\rho_q^2\,.
\end{equation}
With this definition, the kinetic and pairing energies both diverge
but the sum, and indeed the total energy, does not~\cite{[Bul01c]}.
If the pairing energy density involves kinetic or gradient
terms (like it was implemented for the Skyrme interaction SkP~\cite{[Dob84]}),
the regularization would
require the subtraction of higher order terms. This possibility
in not implemented in the present version of the code.
Finally, let us note that
relations (21)--(26) in Ref.~\cite{[Bul01c]} imply that the
variation of quantities $k_F$ and $k_c$ with respect
to the densities should be included in the mean fields.
This contribution is supposed to be small
and is not taken into account in the code. On the other hand,
for density-dependent pairing interaction, the variation with
respect to the explicit density dependence is taken into account
in the code, and gives a contribution to the energy of the order
of a few keV.

\section{The effective Skyrme force}
\label{ch:force}

For a general overview of the foundations and properties of the
Skyrme force we refer the reader to the review article
\cite{[Ben03]} and references therein.
The Skyrme force is an {\em effective} interaction
depending on a limited number of parameters,
\begin{equation}
\begin{split}
V_{12}=&t_0(1+x_0P_\sigma)\delta
+\hf t_1(1+x_1P_\sigma)\left({\mathbf k'}^2\delta+\delta\mathbf k^2\right)
+t_2(1+x_2 P_\sigma)\mathbf k'\cdot\delta\mathbf k \hfill\\
 & +\frac{1}{6}t_3(1+x_3 P_\sigma)\rho^\gamma\delta+\mathrm{i}W_0
\left(\boldsymbol\sigma_1+\boldsymbol\sigma_2\right)\cdot\left(
\mathbf k'\times\delta\mathbf k\right)\,,\hfill
\end{split}
\label{eq:skyrmeforce}
\end{equation}
where $\delta$ is a short notation for $\delta(\bfr_1-\bfr_2)$,
$\mathbf k=\mathbf k_1-\mathbf k_2$ acting on the right and
$\mathbf k'=\mathbf k_1-\mathbf k_2$ acting on the left.

The parameters of the Skyrme forces were fitted in the literature
to reproduce various bulk nuclear properties as well as selected
properties of some nuclei (usually
doubly magic nuclei). Simplifications have often been
made in the expression of the functional~(\ref{skyrme:e}), like
treatment of the Coulomb
exchange term with the Slater approximation and/or omission
of the two-body center of mass contribution or of the ``$\,J^2\,$''
terms.
The latter corresponds to the fourth line in Eq.~(\ref{skyrme:e})
and to the two first terms in the spin-orbit mean field, Eq.~(\ref{mf:so}).
It is important to keep in mind that one should use each parametrization
of the functional within the same simplifications with which it has
been adjusted to data.

For all forces implemented in {\tt HFBRAD}, the
force in the particle-particle channel is chosen to be
\begin{equation}
V_{12}'=\left(t_0'
+\frac{t_3'}{6}\rho^{\gamma'}\right)\delta\,.
\label{eq:skyrmeforcepp}
\end{equation}
Table~\ref{tab:paramph} gives the different sets of parameters
for the force in the particle-hole channel, while table~\ref{tab:parampp}
gives the parameters in the particle-particle channel. It is
important to keep in mind that these latter parameters have been
adjusted along with a given cut-off and cut-off diffuseness, which
are two additional parameters of the force.

\begin{table}[htbp]
  \begin{center}
    \begin{tabular}{|c| *{5}{ r @{.} l }|}
\cline{2-11}
     \multicolumn{1}{c}{} & \multicolumn{2}{|c}{SIII~\protect\cite{[Bei75]}}
                          & \multicolumn{2} {c}{SkM*~\protect\cite{[Bar82]}}
                          & \multicolumn{2} {c}{SLy4~\protect\cite{[Cha98]}}
                          & \multicolumn{2} {c}{SLy5~\protect\cite{[Cha98]}}
                          & \multicolumn{2}{c|}{SkP~\protect\cite{[Dob84]}} \\
\hline
  $t_0$     &$-$1128&75 &$-$2645&0  &$-$2488&913  &$-$2488&345 &$-$2931&6960 \\
  $t_1$     &    395&0  &    410&0  &    486&818  &    484&230 &    320&6182 \\
  $t_2$     &  $-$95&0  & $-$135&0  & $-$546&395  & $-$556&690 & $-$337&4091 \\
  $t_3$     &  14000&0  &  15595&0  &  13777&0    &  13757&0   &  18708&96   \\
  $x_0$     &      0&45 &      0&09 &      0&834  &      0&776 &      0&2921515 \\
  $x_1$     &      0&0  &      0&0  &   $-$0&344  &   $-$0&317 &      0&6531765 \\
  $x_2$     &      0&0  &      0&0  &   $-$1&000  &   $-$1&000 &   $-$0&5373230 \\
  $x_3$     &      1&0  &      0&0  &      1&354  &      1&263 &      0&1810269 \\
  $\gamma$  & 1&0
            & \multicolumn{1}{r@{/}}{1} & \multicolumn{1}{l}{\!\!\!6}
            & \multicolumn{1}{r@{/}}{1} & \multicolumn{1}{l}{\!\!\!6}
            & \multicolumn{1}{r@{/}}{1} & \multicolumn{1}{l}{\!\!\!6}
            & \multicolumn{1}{r@{/}}{1} & \multicolumn{1}{l|}{\!\!\!6}  \\
  $W_0$     &    130&0 &   120&0  &    123&0    &    125&00 &    100&000 \\
\hline
  $J^2$     & \multicolumn{2}{|c}{No} & \multicolumn{2}{c}{No} &
              \multicolumn{2}{c}{No} & \multicolumn{2}{c}{Yes} &
              \multicolumn{2}{c|}{Yes} \\
\hline
    \end{tabular}
\caption{Parameters in the particle-hole channel for
         the different versions of the Skyrme forces implemented
         in the code {\tt HFBRAD}. The last line indicates if the
         $J^2$ terms are included in the Skyrme functional
         when the force is used.\label{tab:paramph}}
  \end{center}
\end{table}

Three kinds of pairing forces have been adjusted: (i) the volume
pairing (the pairing field follows the shape of the density), below
denoted by ``$\rho$'', (ii) the surface pairing (the pairing field is
peaked at the surface and follows roughly the variations of the
density), below denoted by `'$\delta\rho$'', and (iii) the mixed
pairing (a compromise between the two first two), below denoted by
``$\rho$+$\delta\rho$''. All pairing parameters have been adjusted in
order to give a mean neutron gap of 1.245~MeV in $^{120}$Sn.

\begin{table}[htbp]
  \begin{center}
    \begin{tabular}{|l|rrr|rrr|}
\cline{2-7}
      \multicolumn{1}{c|}{} & \multicolumn{3}{c|}{cut-off}
                            & \multicolumn{3}{c|}{regularization} \\[1mm]
\cline{2-7}
      \multicolumn{1}{c|}{} &
      \multicolumn{1}{c}{\ \ \ \ \ $\rho$\ \ \ \ \ } &
      \multicolumn{1}{c}{$\delta\rho$} &
      \multicolumn{1}{c|}{$\rho+\delta\rho$} &
      \multicolumn{1}{c}{\ \ \ \ $\rho\,^r$\ \ \ \ } &
      \multicolumn{1}{c}{$\delta\rho\,^r$} &
      \multicolumn{1}{c|}{$(\rho+\delta\rho)^r$}
 \\[1mm]
\hline
$t_0'$(SIII) &$-$159.6 &$-$483.2 &$-$248.5 &$-$197.6 &$-$807.0 &$-$316.9 \\[1mm]
\hline
$t_0'$(SkM*) &$-$148.6 &$-$452.6 &$-$233.9 &$-$184.7 &$-$798.4 &$-$300.7 \\[1mm]
\hline
$t_0'$(SLy4) &$-$186.5 &$-$509.6 &$-$283.3 &$-$233.0 &$-$914.2 &$-$370.2 \\[1mm]
\hline
$t_0'$(SLy5) &$-$179.9 &$-$504.9 &$-$275.8 &$-$222.7 &$-$901.9 &$-$356.6 \\[1mm]
\hline
$t_0'$(SkP)  &$-$131.6 &$-$429.5 &$-$213.1 &$-$196.6 &$-$1023.0 &$-$326.5  \\[1mm]
\hline
$t_3'$       &       0 & $-37.5\,t_0'$ & $-18.75\,t_0'$
             &       0 & $-37.5\,t_0'$ & $-18.75\,t_0'$ \\[1mm]
\hline
$\gamma'$    &       1 & 1 & 1 & 1 & 1 & 1 \\[1mm]
\hline
    \end{tabular}
\caption{Parameters in the particle-particle channel for
         the different versions of the Skyrme forces implemented
         in the code {\tt HFBRAD}. The three first columns corresponds
         to the use of a cut-off
         at 60~MeV with a Fermi shape and a diffuseness of
         1~MeV, the three columns on the right corresponds to
         the regularization procedure of the pairing field
         described in Sec.~\protect\ref{sec2.6}.
         \label{tab:parampp}}
  \end{center}
\end{table}

\section{Observables and single particle properties}
\label{sec4}

\subsection{Hartree-Fock equivalent energies, radii, and nodes of
quasiparticle wave functions}
\label{sec4:1}

For each quasiparticle state, in addition to its quasiparticle energy
 $E_i$ given by the HFB equation, one has several other
characteristics, which are defined
the following way:
\begin{itemize}
\item the occupation factor $N_i$ which is the norm of the
lower component of the HFB quasiparticle wave function;
\item the (Hartree-Fock) equivalent energy $\bar\varepsilon_i$ which
is defined by applying the BCS type formula to the occupation factor $N_i$, i.e.,
\begin{equation}
N_i=\hf\left(1-\frac{\bar\varepsilon_i-\lambda}{E_i}\right)\,,
\end{equation}
which gives
\begin{equation}
\bar\varepsilon_i=\lambda+E_i\left(2N_i-1\right)\,;
\label{eq:enequiv}
\end{equation}
\item in the same spirit, the equivalent gap $\bar\Delta_i$
is defined by applying the BCS type formula to the
HFB quasiparticle energy $E_i$, i.e.,
\begin{equation}
E_i=\sqrt{(\bar\varepsilon_i-\lambda)^2+\bar\Delta_i^2}\,,
\label{eq:eqpeq}
\end{equation}
which gives
\begin{equation}
\Delta_i=2E_i\sqrt{N_i(1-N_i)}\,;
\label{eq:eqdeq}
\end{equation}
\item the rms radius $\sqrt{\langle r_i^2\rangle}$
is calculated using the lower
component of the quasiparticle wave function
\begin{equation}
\langle r_i^2\rangle=\int_0^{R_{\mathrm{box}}}\!\!
\mathbf r^2\varphi_{2,i}^2(\mathbf r)d^3\mathbf r\,;
\end{equation}
\item the number of nodes of the HFB quasiparticle wave function is defined
as the number of nodes (including the node at the origin but not
the one at infinity) of the component which has the greatest
amplitude.
\end{itemize}
These characteristics of the quasiparticle states can be found
in the output file {\tt hfb\_n\_p.spe}, see Sec.~\ref{output}.

\subsection{Canonical basis}

The canonical states are obtained by diagonalizing the density
matrix (see e.g.\ Refs.~\cite{[RS80],[Dob96]}
for the interpretation of the canonical basis)
\begin{equation}
\int\!d^3\mathbf r'\rho(\bfr,\bfr')\psi_i(\bfr')=v_i^2\psi_i(\bfr)\,.
\end{equation}
As discussed in~Refs.\ \cite{[Dob84],[Dob96]}, all canonical states have
localized wave functions and form a basis. The energies of
the canonical states are defined as the diagonal matrix elements
of the Hartree-Fock field $h$ in the canonical basis,
\begin{equation}
\varepsilon_i=\langle\psi_i|h|\psi_i\rangle ,
\end{equation}
and the pairing gaps associated with these states are the
diagonal matrix elements of the pairing field,
\begin{equation}
\Delta_i=\langle\psi_i|\tilde h|\psi_i\rangle\,.
\end{equation}
Finally, the canonical quasiparticle energy can be defined using the same kind of formula
as~(\ref{eq:eqpeq}),
\begin{equation}
E^{\text{can}}_i=\sqrt{(\varepsilon_i-\lambda)^2+\Delta_i^2}\,,
\label{eq:ecaneq}
\end{equation}
while the occupation probabilities of canonical states
are given by
\begin{equation}
v_i^2=\hf\left(1-\frac{\varepsilon_i-\lambda}{E^{\text{can}}_i}\right)\,,
\label{eq:vcaneq}
\end{equation}
These characteristics of the canonical states can be found
in the output file {\tt hfb\_n\_p.spe}, see Sec.~\ref{output}.

\subsection{Observables and other characteristic quantities of the system}

The rms radii for protons and neutrons are defined as
($q=n$ or $p$)
\begin{equation}
\langle r_q^2\rangle=\int_0^{R_{\mathrm{box}}}\!\!
\mathbf r^2\rho_q^2(\mathbf r)d^3\mathbf r\,.
\label{eq:rnp}
\end{equation}
The charge radius is obtained from the proton radius by taking into
account the proton charge distribution in an approximate way, i.e.,
\begin{equation}
\langle r_{\mathrm{ch}}^2\rangle=\langle r_p^2\rangle+\langle r\rangle_P^2\,.
\label{eq:rcharge}
\end{equation}
with $\langle r\rangle_P=0.8$~fm. The mean gaps are the average
values of the pairing fields
\begin{equation}
\langle\Delta_q\rangle=-\frac{\mathrm{Tr}\,{\tilde h}_q\rho}{A}\,.
\end{equation}
The fluctuations of the particle numbers are defined as
$\langle \hat N_q^2-\langle\hat N_q\rangle^2\rangle$ and
given by $2\mathrm{Tr}\left[\rho_q^2-\rho_q\right]$.

Finally, the
rearrangement energy, which comes from the density dependence of
the force, and which shows how much the force is modified by the medium
effects, is given by
\begin{eqnarray}
E_{\mathrm{rear}}&=&\dps
  \frac{t_3}{24}\gamma\rho^\gamma\left[\left(
  1+\frac{x_3}{2}\right)\rho^2-\left(x_3+\hf\right)
  \sum_q\rho_q^2\right]\hfill\nonumber\\
 &+&\dps
  \frac{t_3'}{48}\gamma'(1-x_3')\rho^{\gamma'}\sum_q
  \tilde\rho_q^2
  \,-\,\frac{e^2}{4}\left(\frac{3}{\pi}\right)^{\frac{1}{3}}\rho_p^{4/3}\,.
\end{eqnarray}

\section{Numerical treatment of the problem}
\label{sec5}

\subsection{The problem inside a box}

The system of differential equations~(\ref{HFBdiff}) is solved
in coordinate representation in a spherical box for a given
choice of boundary conditions. Here we briefly introduce notations that
useful for the discussion that follows and discuss the validity
of approximations to solve the differential problem.
The radial coordinate $r$ is approximated by a mesh of points spaced by the step $h$:
$r_n = n h$, with $n \in \{0,1,...,N\}$.
For any quantity $F(r)$ approximated on the mesh we use notation
$F_n\equiv F(r_n)$.

On the wall of the box, we use two possible boundary conditions:
vanishing wave functions (Dirichlet condition) or vanishing derivatives
of the wave functions
(Neumann condition). With this choice, the continuous part
of the spectrum is discretized. Important questions are
how well the continuum is approximated by this discretization
and whether it does not introduce numerical artifacts when solving
system of equations~(\ref{HFBdiff}).
A discussion concerning the behaviour of the resonant like
solution inside the box can be found in~\cite{[Dob84]}.
The non-linearity of the equations
makes it difficult to discuss further the effect of the
discretization of the spectrum, nevertheless, the
numerical examples shown in Sec.~\ref{expl} show
that such numerical method provides a fairly good
approximation of the exact problem.
In particular, it can be observed that once a sufficiently large
box is used, the global properties of the nucleus become
perfectly independent of its size.

\subsection{The Numerov algorithm for the HFB equation}
\label{sec:numerov}

The Numerov (Cowell) method \cite{[Dah74]} is the standard
technique for a precise integration of second-order differential
equations; here we only give some details pertaining to
its application to the system of two HFB equations (\ref{HFBdiff}).
Let function $y(r)$ be the solution of the differential equation
\begin{equation}
y''=F(y,r)\,.
\label{eq:eqdif}
\end{equation}
The Numerov algorithm is based on the finite difference formula
for three consecutive points on a mesh with step $h$,
\begin{equation}
y_{n+1}-2 y_n+y_{n-1}=
  \frac{h^2}{12}\left(y''_{n-1}+10 y''_n+y''_{n+1}\right)\,+\,\po({h^6}),
\label{eq:numerova}
\end{equation}
where $\po({h^6})$ means that the error in this relation is of the
order $h^6$. Combining this formula with~(\ref{eq:eqdif}), one obtains
the relation
\begin{equation}
\left(1-\frac{h^2}{12}F_{n+1}\right)y_{n+1}
-2\left(1+\frac{5h^2}{12}F_n\right)y_n
+\left(1-\frac{h^2}{12}F_{n-1}\right)y_{n-1}=0 ,
\label{eq:numerov}
\end{equation}
which is obviously also valid if $y$ is a vector and $F$ a square
matrix. Introducing notations
\begin{equation}
A_n=\left(1-\frac{h^2}{12}F_n\right),
\end{equation}
and
\begin{equation}
G_n=A_n y_n\,,
\label{eq:Gn}
\end{equation}
one obtains the recurrence relation (Numerov iterations),
\begin{equation}
\left\{\begin{matrix}
G_{n+1} =& 12  y_n - 10  G_n - G_{n-1}\,, \hfill\\[1.5mm]
\hfill y_{n+1} =& A_{n+1}^{-1}G_{n+1}\,. \hfill\\
\end{matrix}\right.
\label{eq:rec}
\end{equation}
In the present applications, $A_n$ is a $2\times2$ matrix
(position and energy dependent) and
the calculation of its inverse could be the bottleneck of
the procedure if it is not computed cleverly. This can be done
in the following way. Using the
notations introduced in Sec.~\ref{sec:hfbeq}, one can write
matrix $A_n$ in an explicit form as
\begin{equation}
A_n=\left(\begin{matrix}
1 - \frac{h^2}{12}\frac{V_n}{M^*_n}+\frac{h^2}{12}\frac{E}{M^*_n}
    & -\frac{h^2}{12}\frac{W_n}{M^*_n} \\[1.5mm]
\frac{h^2}{12}\frac{W_n}{M^*_n}
    & 1 - \frac{h^2}{12}\frac{V_n}{M^*_n}-\frac{h^2}{12}\frac{E}{M^*_n} \\
\end{matrix}\right) ,
\end{equation}
whereby its inverse is given by
\begin{multline}
\dps A^{-1}_n=\frac{1}%
{\left(1 - \frac{h^2 V_n}{12 M^*_n}\right)^2+
\left(\frac{h^2 W_n}{12 M^*_n}\right)^2-E^2
\left(\frac{h^2}{12 M^*_n}\right)^2} \nonumber\\
\dps\hskip 4cm\left(\begin{matrix}
1 - \frac{h^2}{12}\frac{V_n}{M^*_n}-\frac{h^2}{12}\frac{E}{M^*_n}
    & \frac{h^2}{12}\frac{W_n}{M^*_n} \\[1.5mm]
-\frac{h^2}{12}\frac{W_n}{M^*_n}
    & 1 - \frac{h^2}{12}\frac{V_n}{M^*_n}+\frac{h^2}{12}\frac{E}{M^*_n} \\
\end{matrix}\right)\,.
\end{multline}
It appears clearly that a lot of time can be saved by calculating
and storing
the energy independent terms,
\begin{equation}
1 - \frac{h^2 V_n}{12 M^*_n}\,,
\ \frac{h^2}{12 M^*_n}\,, \ \frac{h^2 W_n}{12 M^*_n}
\ \ \mbox{and}
\ \ \left(1 - \frac{h^2 V_n}{12 M^*_n}\right)^2+
\left(\frac{h^2 W_n}{12 M^*_n}\right)^2 ,
\end{equation}
only once in each $(\ell j q)$-block; then the energy-dependent terms
can be in each iteration calculated very rapidly.

The eigen energies are found by integrating two linearly
independent solutions, $y^{(1+)}(r)$ and $y^{(2+)}(r)$,
from the origin to a given point $r_m$, called the matching point.
These two solutions are selected by imposing that either
the first or the second component of the quasiparticle wave function
vanishes near the origin, i.e.,
\begin{equation}
y^{(1+)}(r)\sim\left(\begin{matrix} r^{\ell+1}\\ 0\\
             \end{matrix}\right)\,,
\ \ \ %
y^{(2+)}(r)\sim\left(\begin{matrix} 0 \\ r^{\ell+1}\\
             \end{matrix}\right)\,,
\ \ \ \mbox{for}\ r\rightarrow 0\,,
\label{eq:bounda}
\end{equation}
whereby we have
\begin{equation}
y^{(1+)}_0 = \left(\begin{matrix} 0\\ 0\\
             \end{matrix}\right)\,,
\ \ \ %
y^{(2+)}_0 = \left(\begin{matrix} 0 \\ 0\\
             \end{matrix}\right)\,,
\label{eq:bounda0}
\end{equation}
and
\begin{equation}
y^{(1+)}_1 = \left(\begin{matrix} Ah^{\ell+1}\\ 0\\
             \end{matrix}\right)\,,
\ \ \ %
y^{(2+)}_1 = \left(\begin{matrix} 0 \\ Bh^{\ell+1}\\
             \end{matrix}\right)\,.
\label{eq:bounda1}
\end{equation}

Next, another two linearly
independent solutions, $y^{(1-)}(r)$ and $y^{(2-)}(r)$, are found
by the backward integration from the wall of the box $R_{\mathrm{box}}\equiv r_N$
to $r_m$, again imposing that either
the first or the second component of the quasiparticle wave function
vanishes near $r_N$ for the Dirichlet boundary conditions, i.e.,
\begin{equation}
y^{(1-)} \sim \left(\begin{matrix} r-r_N\\ 0\\
             \end{matrix}\right)\,,
\ \ \ %
y^{(2-)} \sim \left(\begin{matrix} 0 \\ r-r_N\\
             \end{matrix}\right)\,,
\ \ \ \mbox{for}\ r\rightarrow r_N\,,
\label{eq:boundb}
\end{equation}
whereby we have
\begin{equation}
y^{(1-)}_N = \left(\begin{matrix} 0\\ 0\\
             \end{matrix}\right)\,,
\ \ \ %
y^{(2-)}_N = \left(\begin{matrix} 0 \\ 0\\
             \end{matrix}\right)\,,
\label{eq:boundb0}
\end{equation}
and
\begin{equation}
y^{(1-)}_{N-1} = \left(\begin{matrix} Ch\\ 0\\
             \end{matrix}\right)\,,
\ \ \ %
y^{(2-)}_{N-1} = \left(\begin{matrix} 0 \\ Dh\\
             \end{matrix}\right)\,.
\label{eq:boundb1}
\end{equation}
For the Neumann boundary conditions, at the left-hand-sides
of Eqs.~(\ref{eq:boundb})--(\ref{eq:boundb1}) one replaces
functions by derivatives.
Finally, one obtains four solutions valid everywhere
except at the matching point $r_m$ and depending on
the four constants $A$, $B$, $C$, and $D$.

The boundary conditions~(\ref{eq:bounda0})--(\ref{eq:bounda1})
and~(\ref{eq:boundb0})--(\ref{eq:boundb1})
determine the initial values $(G_0, G_1)$ and
$(G_N, G_{N-1})$, respectively, needed to start the Numerov iterations of~(\ref{eq:rec}).
Obviously, a correct limit should be taken when calculating
values of $G_0$ from Eq.~(\ref{eq:Gn}). It turns out that one
obtains $G_0$=0 for all partial waves except $\ell=1$, which is the
case that deserves a little more attention. For example,
focusing our attention on one of the solutions,
\begin{equation}
y^{(1+)}(r)=\left(\begin{matrix} Ar^2\\ 0\\
             \end{matrix}\right)\,,
\end{equation}
we use the fact that in the vicinity of the origin $A(r)$ is dominated by the centrifugal
term, so that its diagonal matrix elements are close to
$-\frac{h^2}{12}\frac{2}{r^2}$.
This leads to the result
\begin{equation}
G^{(1+)}_0=\lim_{r\rightarrow 0}A(r)y(r)= - \frac{Ah^2}{6}
\left(\begin{matrix} 1 \\ 0 \\ \end{matrix}\right).
\end{equation}
An analogous result can be obtained for the other solution at the origin.

The eigenenergies are found by requiring that the full
solution
$y=A y^{(+)}_1 + B y^{(+)}_2 = C y^{(-)}_1 + D y^{(-)}_2$
and its first derivative are continuous at $r=r_m$
(see for example \cite{[Bel87]}).
This matching conditions read
\begin{equation}
\left(\begin{matrix}
y_m^{(1+)} & y_m^{(2+)} & - y_m^{(1-)} & - y_m^{(2-)} \\
{y'}_m^{(1+)} & {y'}_m^{(2+)} & - {y'}_m^{(1-)} & - {y'}_m^{(2-)} \\
\end{matrix}\right)
\left(\begin{matrix}
A \\ B \\ C \\ D \\
\end{matrix}\right)=M_4
\left(\begin{matrix}
A \\ B \\ C \\ D \\
\end{matrix}\right)=0\,.
\label{eq:matching}
\end{equation}
The HFB quasiparticle wave functions
having two components, the matching conditions amount
to searching for zeros of the determinant of the $4\times4$ matrix $M_4$
in function of the quasiparticle energy.
For $\ell=0$, the matching point $r_m$ is chosen to be near the point where the central
part of the Hartree-Fock potential is half of its maximum,
while for increasing values of $\ell$ it is
gradually shifted to the exterior.  A systematic search allows to find
the intervals where the zeros of this determinant are located, and
then the Newton method
is used to find their exact locations to an arbitrary precision.

For vanishing determinant, Eq.~(\ref{eq:matching}) is used
to determine three of the multiplicative
constants in terms of the fourth one.
This is done by extracting a $3\times3$ matrix $M_3$ out
of $M_4$ and inverting it.
Among the four possible choices one may have for $M_3$ we retain
the one for which the product $M_3\times M_3^{-1}$ is
closest to unity. Finally, the fourth multiplicative
constant is fixed by the normalization condition of the wave
function.

When approaching the self-consistent solution, the eigenenergies do
not change very much between two consecutive Bogolyubov iterations. We
take advantage of this fact by keeping in memory the intervals where
the solutions are located and using them as initial guesses for the
next iteration. This results in finding solutions for eigenenergies
faster and faster as we get closer to the converged self-consistent
state.

\subsection{Numerical instabilities}
\label{sec:instab}

If we try to solve the HFB equations in rather big boxes
(typically beyond 30 or 40~fm depending on the mass of the
nucleus) we encounter strong numerical instabilities.
This problem is not trivially related to the accumulation
of errors during the Numerov iteration, because
it disappears if the nucleus is treated within the HF
approximation instead of HFB. The origin of the problem is in the fact
that at large distances the upper and lower HFB wave functions
differ by many orders of magnitude, and thus calculations
within any number of significant digits must fail at
some distance, especially if the small component
decreases very fast exponentially. This is so because the small component becomes
easily polluted by the large component even if they are coupled
by very small matrix elements.

This numerical instability
can be removed by neglecting the pairing
field beyond a certain large distance $R_{\mathrm{cut}}$. Instead of the
full pairing field we consider the truncated field
given by
\begin{equation}
\label{Rcut}
\Delta_{\mathrm{cut}}(r)=
\left\{
  \begin{matrix}
    \Delta(r) \hfill & \mbox{for\ } r \le R_{\mathrm{cut}}\,, \hfill \\[1mm]
    0 \hfill         & \mbox{for\ } r > R_{\mathrm{cut}}\,. \hfill\\
\end{matrix}\right.
\end{equation}
It will be shown in section~\ref{par:exa} that this truncation
has no significant effects on the solution if $R_{\mathrm{cut}}$ is
far enough from the origin. Note that by removing the coupling between
the lower and upper components one also removes the second term in
the asymptotic form of the lower component, see Eq.~(\ref{eq:asympt}).

Another source of numerical instabilities may come from an accidental
degeneracy of deep hole and particle states. In the HFB
approach the deep hole states are embedded in the
continuum of particle states. The deep hole states are extremely narrow and thus
insensitive to the size of the integration box. On the
other hand, the discretized particle states are driven by the boundary
condition and their spectrum becomes denser as the box
is enlarged. As a result, a deep hole state and a quasiparticle
(particle like) state can become quasi degenerate and the
systematic search for the solution [corresponding to the matching
condition (\ref{eq:matching})] can miss both of them simultaneously. This situation can
show up during the iterations and disappear at the final solution; in such a
case this numerical accident has no consequence on the final
solution. If the quasi degeneracy is met close to the final
self-consistent solution, the iterations may not converge.
This can be overcome by using a smaller energy step
or by slightly changing the size of the integration box
in order to lift the degeneracy.

\section{Input data file}
\label{input}

The code {\tt HFBRAD} is driven using a single input file
which contains all information on the geometry of the system,
parameters of the force, demanded output and
nuclei to be computed.
The data are read in namelist, which is a Fortran-90/95 feature.
All variables in a namelist are optional and acquire a default
value when absent.

\subsection{The namelist {\tt input}}

The namelist input contains the variables summarized on
table~\ref{tab:input} with their default values. The variables
are
\begin{itemize}
\item {\tt force}: sets the parameters of the force in the
particle-hole channel, possible values are SIII, SKM*, SKP,
SLY4 and SLY5,
it must be written in uppercase.
\item {\tt mesh\_points}: number of mesh points in the
box minus one since integration is
made from $i=0$ to $i=\mbox{\tt{mesh\_points}}$.
\item {\tt integ\_step}: distance between two points
on the mesh, in fermis.
\item {\tt itmax}: maximum number of iterations to obtain
the convergence of the mean fields and total energy.
\item  {\tt bogolyubov(2)}: logical variables to choose between
the HF or HFB approximation, the first element is for neutrons
and the second one for protons.
\item {\tt eps\_energy}: maximum relative variation on the
total energy before the iterations stops.
\item {\tt max\_delta}: in addition to the previous convergence
condition, the maximum variation on the sum of the neutron
and proton pairing gaps must be smaller than this value.
\item {\tt regularization}: logical variable used to
switch on or off the regularization of the abnormal density.
\item {\tt pairing\_force}: integer used to choose the pairing
interaction between: full Skyrme force (0), volume pairing (1),
surface pairing (2) or mixed pairing (3).
\item {\tt boundary\_condition}: integer, Dirichlet condition
at the box radius (0), Neumann (1), Dirichlet for even $\ell$
and Neumann for odd $\ell$ (2), or Neumann for even $\ell$
and Dirichlet for odd $\ell$ (3).
\item {\tt xmu}: part of the field of the previous iteration
which is kept to build the field for the next iteration.
\end{itemize}

\begin{table}[htbp]
 \begin{center}
  \begin{tabular}{|l|c|r|}
\hline
   Variable & type & Default value \\
\hline
   {\tt force}                & character*4 & {\tt   SLY4}  \\
   {\tt mesh\_points}         & integer     & {\tt     80}  \\
   {\tt integ\_step}          & real        & {\tt   0.25}  \\
   {\tt itmax}                & integer     & {\tt    680}  \\
   {\tt bogolyubov(2)}        & logical     & {\tt (/T,T/)}  \\
   {\tt eps\_energy}          & real        & {\tt  1.e-8}  \\
   {\tt max\_delta}           & real        & {\tt  5.e-7}  \\
   {\tt regularization}       & logical     &      {\tt F}  \\
   {\tt pairing\_force}       & integer     & {\tt      1}  \\
   {\tt boundary\_condition}  & integer     & {\tt      0}  \\
   {\tt xmu}                  & real        & {\tt    0.8}  \\
\hline
  \end{tabular}
  \caption{Variables from the namelist input. The type {\it real} means
           {\it real} of the kind chosen in the module in {\tt cste.f90}.
           \label{tab:input}}
 \end{center}
\end{table}

\subsection{The namelist {\tt nucleus}}

This namelist describes the nucleus to be computed, it can be repeated
as many times as needed in the input file. Along with the neutron
and proton numbers, other quantities can be set, notably the parameters
of the Skyrme force in the particle-particle channel.
If a variable is not present in the list and has not been present
in the previous namelists {\tt nucleus}, it acquires a default
value (see table~\ref{tab:nucleus}). If a variable is not present
but was present in a previous namelist, it keeps its previous value.
The variables are
\begin{itemize}
\item {\tt neutron} and {\tt proton}: number of neutrons and protons
for the nucleus. If one of this number is negative then the program
will let vary the number of particles in order to reach the
corresponding drip line.
\item {\tt j\_max(2)}: maximum values of $2J$ (first for neutrons,
second for protons). It is reasonable to use different values
for neutrons and protons in the case of very asymmetric or
semi magic nuclei.
\item{\tt cut\_off} and {\tt cut\_diffuseness}: If the regularization
procedure is used, the densities are built by summing up contributions from the
quasiparticle states up to the maximum equivalent energy (\ref{eq:enequiv})
$E_{\mathrm{max}}$ given by {\tt cut\_off},
if not, the summation is made up to the cut-off equivalent energy
$E_{\mathrm{cut}}$
(\ref{eq:enequiv}) given by
{\tt cut\_off} with a smooth Fermi profile of diffuseness
{\tt cut\_diffuseness}, see Sec.~\ref{sec2.6}.
\item {\tt r\_cut}: value of distance $R_{\mathrm{cut}}$ beyond which the paring fields
are neglected when integrating the HFB equations, see Eq.\ (\protect\ref{Rcut}).
\item {\tt e\_step}: initial step in energy (in MeV) for the
search of the solution of the HFB equations.
\item {\tt skt0p} and {\tt skt3p}: parameters of the force in
the particle-particle channel as defined in equation~(\ref{eq:skyrmeforcepp}).
\item {\tt read\_pot}: This string can be used to give a name
of a file where the initial potentials are read. The number
of mesh points can be different in the run and in the saved
potentials but the integration step must be the same.
\item {\tt densities}, {\tt meanfields}, {\tt quasiparticle}
and {\tt canonical\_states}: logical flags which can be set
if one wants to save the densities, mean fields and quasiparticle
and canonical states wave functions at the end of the run.
\end{itemize}
\begin{table}[htbp]
 \begin{center}
  \begin{tabular}{|l|c|r|}
  \hline
   Variable          & type         & Default value \\
  \hline
  neutron            & integer      &            -- \\
  proton             & integer      &            -- \\
  j\_max(2)          & integer      &     (/21,21/) \\
  cut\_off           & real         &          60.0 \\
  cut\_diffuseness   & real         &           1.0 \\
  r\_cut             & real         &          30.0 \\
  e\_step            & real         &    calculated \\
  skt0p              & real         &    predefined \\
  skt3p              & real         &    predefined \\
  read\_pot          & character*40 &          ``'' \\
  densities          & logical      &             F \\
  meanfields         & logical      &             F \\
  quasiparticles     & logical      &             F \\
  canonical\_states  & logical      &             F \\
  \hline
  \end{tabular}
\caption{Variables from the namelist {\tt nucleus}. The neutron
         and proton numbers have no default values. If the energy
         step is not present in the list it is calculated as a function
         of the box radius. By default, the values of {\tt skt0p}
         and {\tt skt3p} are fixed by the kind of pairing force
         chosen in the namelist {\tt input}.
         \label{tab:nucleus}}
 \end{center}
\end{table}

\section{Output files} \label{output}

In addition to the optional files containing the densities, mean fields
and wave functions,
at the end of the run the following output files are saved:
\begin{itemize}
\item {\tt hfb\_n\_p.spe}: where {\tt n} and {\tt p} are the neutron
and proton numbers. This file contains the quasiparticle spectrum,
the canonical states spectrum (if the flag {\tt canonical\_states}
was set) and the mean fields.
\item {\tt hfb.summary}: this files summarizes the global
properties for all the nuclei in the run.
\end{itemize}

\section{Examples} \label{expl}

\subsection{Self-consistent calculation for $^{150}$Sn}
\label{par:exa}

As a first example we discuss here the results obtained for
the semi magic isotope $^{150}$Sn ($N=100$ and $Z=50$)
using the effective force SLy4$^{\rho+\delta\rho}$.
This
nucleus is an interesting example since it has a rather small
neutron Fermi energy ($\lambda_N\simeq -1.066$~MeV).
Small neutron separation energy allows us to discuss
possible effects of the integration box size as well as
the choice of the boundary condition.
If not explicitly stated otherwise, in all calculations the
integration step is $h=0.2$~fm
for the box size of $R_{\mathrm{box}}=30$~fm, i.e., the number of
integration points is 150+1.
The global properties
of the nucleus for Dirichlet or Neumann boundary
condition are shown in table~\ref{tab:b}.
\begin{table}
\begin{center}
\begin{tabular}{|l|cccc|}
\hline
Boundary condition & $E$ & $\langle\Delta_N\rangle$ &
    $\lambda_N$ & $\sqrt{\langle r_N^2\rangle}$ \\
\hline
Dirichlet   & -1131.863 & -1.431 & -1.067 & 5.264 \\
Neumann     & -1131.862 & -1.431 & -1.067 & 5.264 \\
\hline
\end{tabular}
\caption{Total binding energy, mean pairing gap, Fermi energy (all in MeV)
         and neutron rms radius (in fm) in $^{150}$Sn, calculated for the force
         SLy4$^{\rho+\delta\rho}$ and two different
         boundary conditions: Dirichlet (vanishing wave functions
         at $r=R_{\mathrm{box}}$) or Neumann (vanishing derivative of
         the wave functions at $r=R_{\mathrm{box}}$).
         \label{tab:b}}
\end{center}
\end{table}

In addition to the small neutron separation energy,
one finds several single particle states with a significant
occupation around $\lambda_N$
with different values of $\ell$ and indeed different asymptotic
bahaviour. The single particle neutron states in the vicinity
of the Fermi energy are shown in table~\ref{tab:a}. It is
gratifying to observe that despite the fact that the quasiparticle
states are a little bit influenced by the change of the boundary
conditions, the canonical states remain unchanged.
\begin{table}
\begin{center}
\begin{tabular}{|l|l|cccc|}
\hline
Boundary condition
&$n\ell j$          &$3p_{1/2}$&$3p_{3/2}$&$2f_{5/2}$&$1h_{9/2}$\\
\hline
Dirichlet
&$E$ (MeV)          &\ 0.926 \ &\ 0.937\  &\ 1.330\  &\ 1.556\  \\[0.5mm]
&$N$                &  0.221   &  0.577   &  0.254   &  0.438   \\[0.5mm]
&$\varepsilon$ (MeV)&  1.169   &  1.051   &  1.445   &  1.572   \\[0.5mm]
&${v}^2$            &  0.251   &  0.619   &  0.264   &  0.438   \\[0.5mm]
\hline
Neumann
&$E$ (MeV)          &  0.919   &  0.932   &  1.328   &  1.556   \\[0.5mm]
&$N$                &  0.211   &  0.562   &  0.251   &  0.438   \\[0.5mm]
&$\varepsilon$ (MeV)&  1.168   &  1.051   &  1.445   &  1.572   \\[0.5mm]
&${v}^2$            &  0.251   &  0.619   &  0.263   &  0.438   \\[0.5mm]
\hline
\end{tabular}
\caption{Single particle neutron levels in $^{150}$Sn using the
         effective force SLy4$^{\rho+\delta\rho}$. Calculations
         have been made in a 30\,fm box with an integration step of
         0.2\,fm. $E$ and $E$ are the quasiparticle energies
         and norms of the lower components while $\varepsilon$ is
         the canonical energy (\ref{eq:ecaneq}) and ${v}^2$ is the occupation
         factor of the corresponding canonical state (\ref{eq:vcaneq}).
         \label{tab:a}}
\end{center}
\end{table}

\begin{figure}[htbp]
\begin{center}
$\begin{matrix}
\includegraphics*[scale=0.51,bb=37 52 392 497]{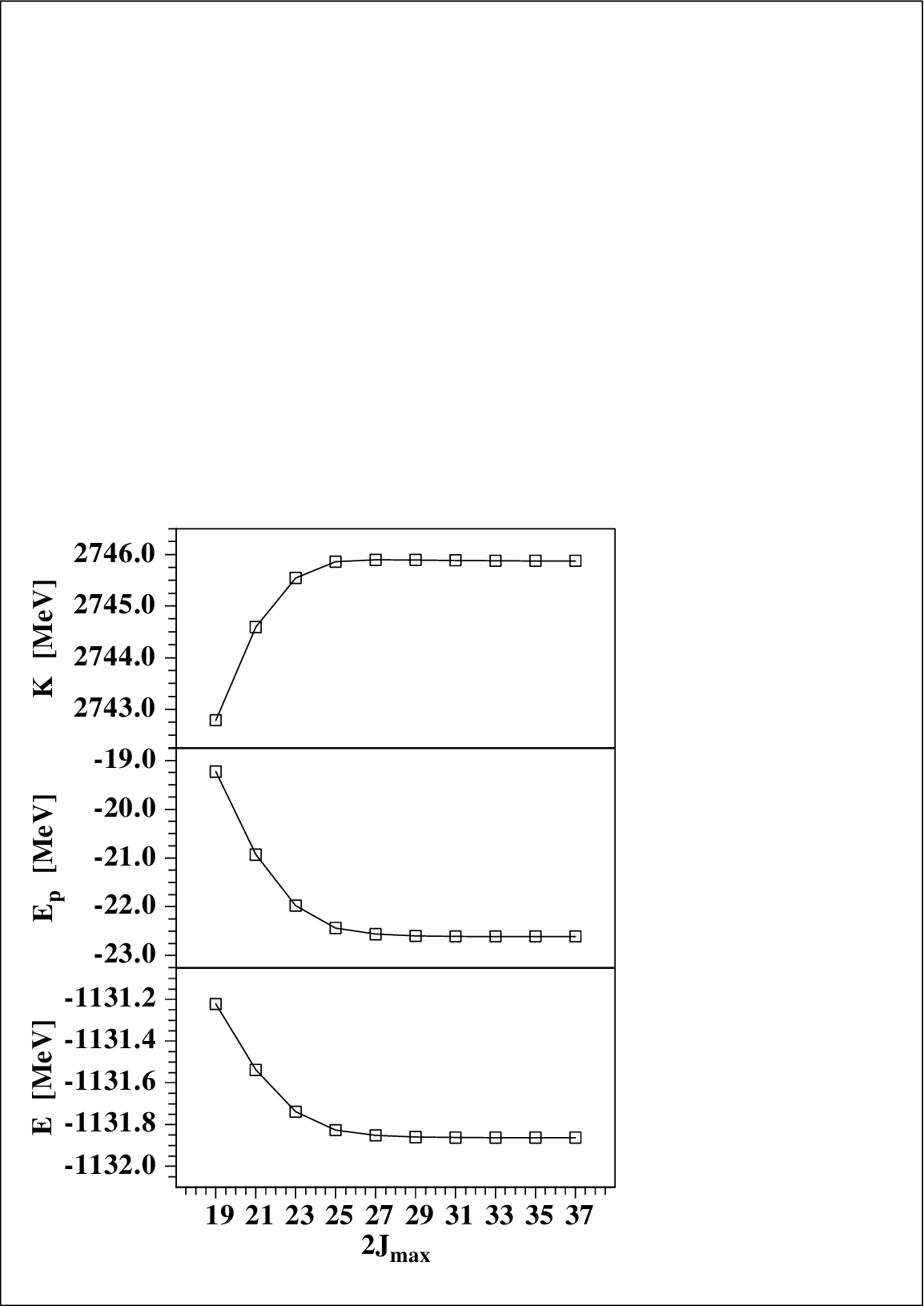} \ \ &
\includegraphics*[scale=0.51,bb=22 52 404 497]{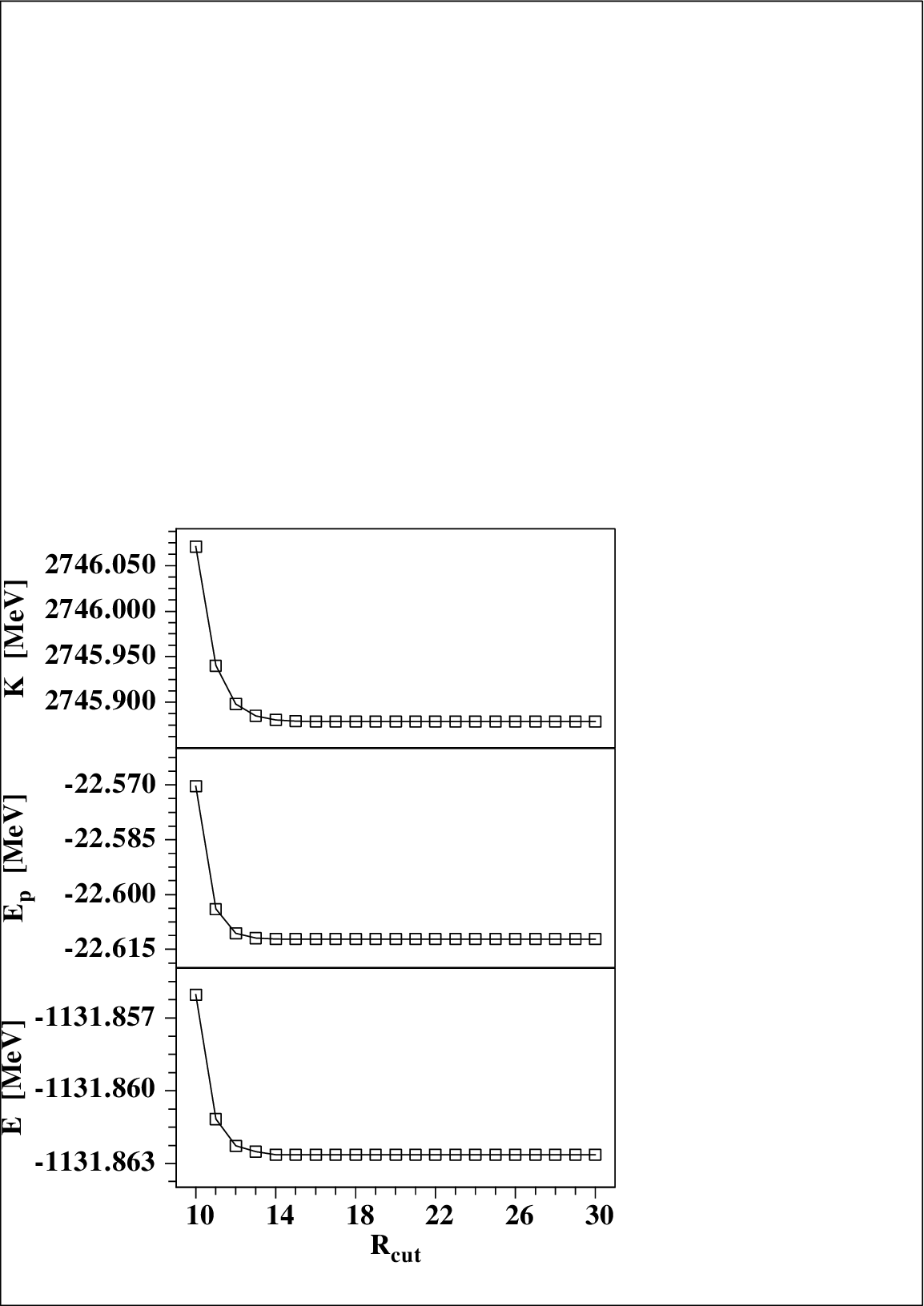} \\
\end{matrix}$
\caption{Kinetic (top panels), pairing (middle panels) and
  total energies (lower panels) for $^{150}$Sn as functions
  of $2J_{\mathrm{max}}$ in a 30~fm box with an integration step of
  0.2~fm (left) and
  of the pairing-field cut-off radius $R_{\mathrm{cut}}$ defined
  in Eq.\ (\protect\ref{Rcut}) (right).
  \label{fig:a}}
\end{center}
\end{figure}

When a rather exotic nucleus is studied, it is important
to check how well the results are converged with respect to
the various approximations which are made in the numerical
treatment of the HFB problem. First of all, we limit the
calculation to a maximum number of partial wave given
by $J\leq J_{\mathrm{max}}$, which has to by consistent with the
mass of the nucleus.
The convergence of the kinetic, pairing,
and total energies are shown in the left part of Fig.~\ref{fig:a}.
Between $2J_{\mathrm{max}}=31$ and 37 the total energy
difference is 1.3\,keV.
The changes in the kinetic and pairing energies are a little
bit bigger (7\,keV and 6\,keV, respectively); since these
quantities are not at their extrema,
they react more rapidly to a small change of the wave function.

As discussed in section~\ref{sec:instab}, the introduction of
the cut-off radius $R_{\mathrm{cut}}$ in the pairing channel,
Eq.~(\ref{Rcut}), can improve
the numerical stability of the iterative procedure when
solving the HFB equation. The right part of Figure~\ref{fig:a}
displays the evolution of the kinetic, pairing, and total
energies with the same parameters as in the left part of the
Figure, except for the value of $2J_{\mathrm{max}}$ which is fixed at 31. It can
be clearly seen that beyond a rather small distance, the total energy
and its components are not affected by the introduction of this
cut-off. Specifically, for $R_{\mathrm{cut}}\geq17$~fm the change in
the total energy is less than 0.001\,keV.

\begin{figure}[htbp]
\begin{center}
\includegraphics*[scale=0.55,bb=74 53 489 345]{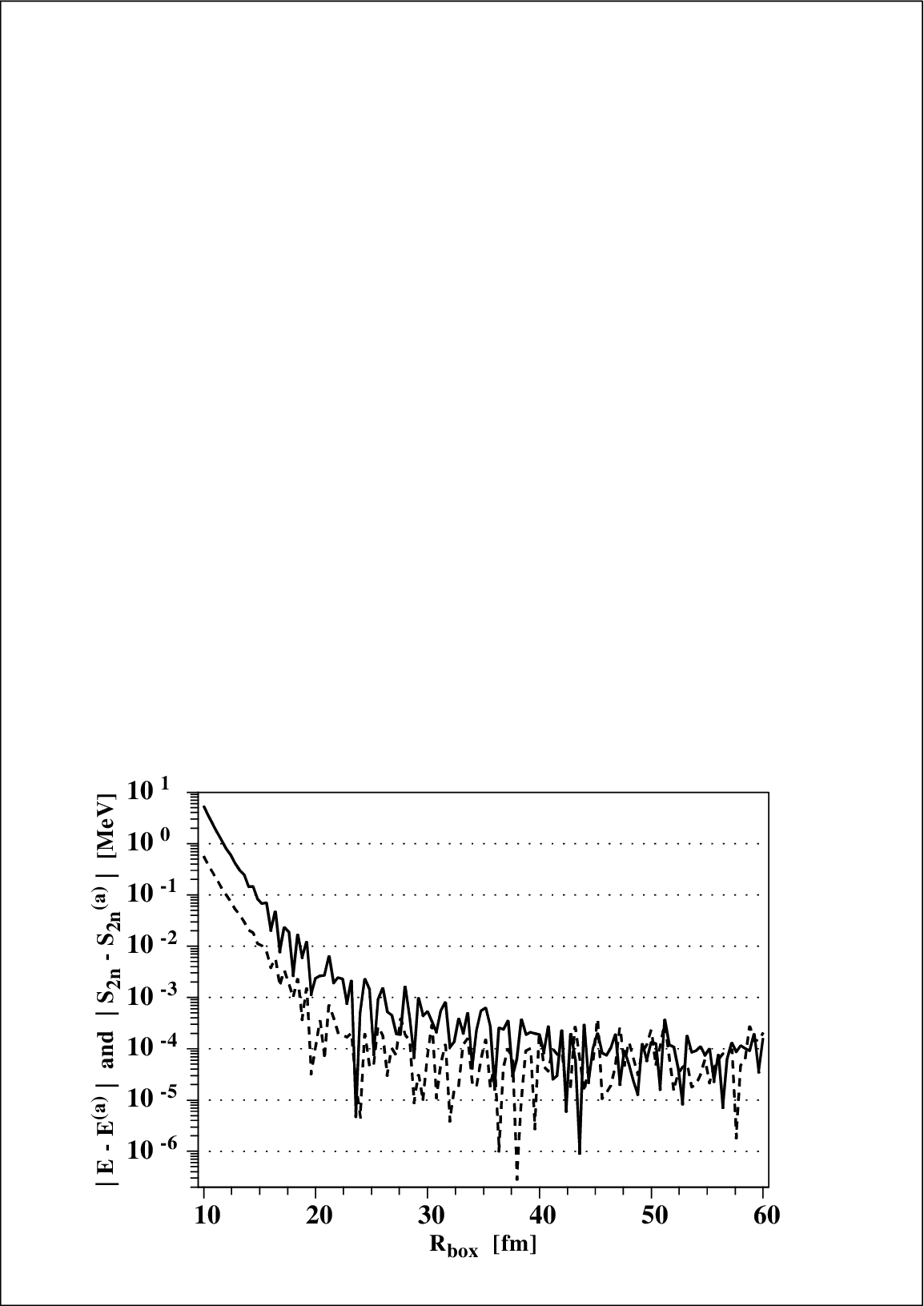}
\caption{Difference between the total
  energy $E$ of $^{150}$Sn (solid line) and its average asymptotic value $E^{(a)}$
  (see text) as a function
  of the box size $R_{\mathrm{box}}$, and the analogous difference for
   the two-neutron separation energy $S_{2n}$ (dashed line).\label{fig:b}}
\end{center}
\end{figure}
Figure~\ref{fig:b} shows the convergence of the total energy $E$
and two-neutron separation energy $S_{2n}$ as functions of the
size of the box. Calculations have been made for $R_{\mathrm{box}}$ ranging
from 10~fm to 60~fm with a step of 0.4~fm. The cut-off equivalent energy of
$E_{\mathrm{cut}}$=60\,MeV and the diffuseness of 1\,MeV have been used.
In order to clearly exhibit the asymptotic trend of these
quantities we have estimated their average asymptotic values
$E^{(a)}$ and $S_{2n}^{(a)}$.
Since for $R_{\mathrm{box}}\geq35$~fm
the results
do not show any significant evolution, these values are
estimated by taking averages
for 35~fm$\leq R_{\mathrm{box}}\leq60$~fm, which gives
$E^{(a)}=-1\,131.862\,114$\,MeV and
$S_{2n}^{(a)}=-1.925\,548$\,MeV.
Here, the last significant digit corresponds to 0.001\,keV, which is the
accuracy required to stop the iterations when we solve the HFB
equations for any given value of $R_{\mathrm{box}}$.
The fluctuations around the average values
result from the individual continuum states entering into
the cut-off window with increasing $R_{\mathrm{box}}$.
We see that for relatively small boxes ($R\leq20$~fm) the
two-neutron separation energy is an order of magnitude more
stable than the total binding energy of the nucleus.
For $R\geq30$~fm, the two quantities reach their asymptotic
values with a random dispersion of about 0.5~keV.
The value of $R_{\mathrm{box}}=60$~fm is obviously unnecessarily
large but it represents a good test of the stability of the
integration procedure. From these results we conclude that
the solution of the HFB equations with the box boundary conditions
and the energy cut-off is precise up to about 1\,keV.

\begin{figure}[htbp]
\begin{center}
\includegraphics*[scale=0.52,bb=67 55 488 326]{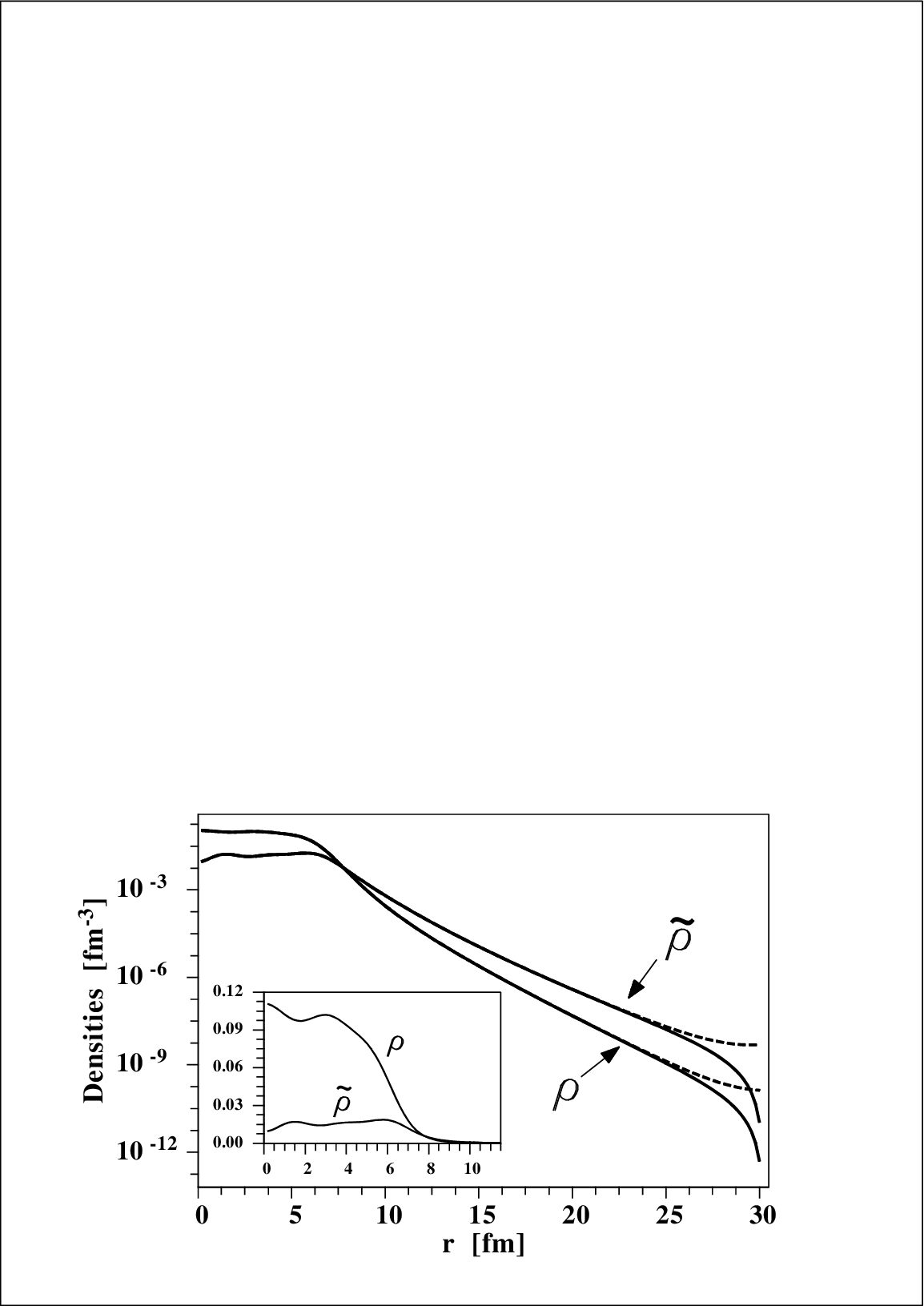}
\caption{Neutron particle and pairing densities in $^{150}$Sn
  for Dirichlet (solid line) and Neumann (dashed line)
  boundary conditions. The inset represents the same
  quantities on a linear scale, in that case the impact of
  the boundary condition cannot be seen.\label{fig:e}}
\end{center}
\end{figure}

A typical asymptotic behaviour of the particle and pairing densities
for neutrons
is shown in Fig.~\ref{fig:e} for two different choices of the boundary condition.
The impact of the boundary conditions only shows
up within about 5\,fm near the box edge. It is
small enough to have no significant effect on the calculated
observables.

\begin{figure}[htbp]
\begin{center}
$\begin{matrix}
\includegraphics*[scale=0.42,bb=36 55 487 328]{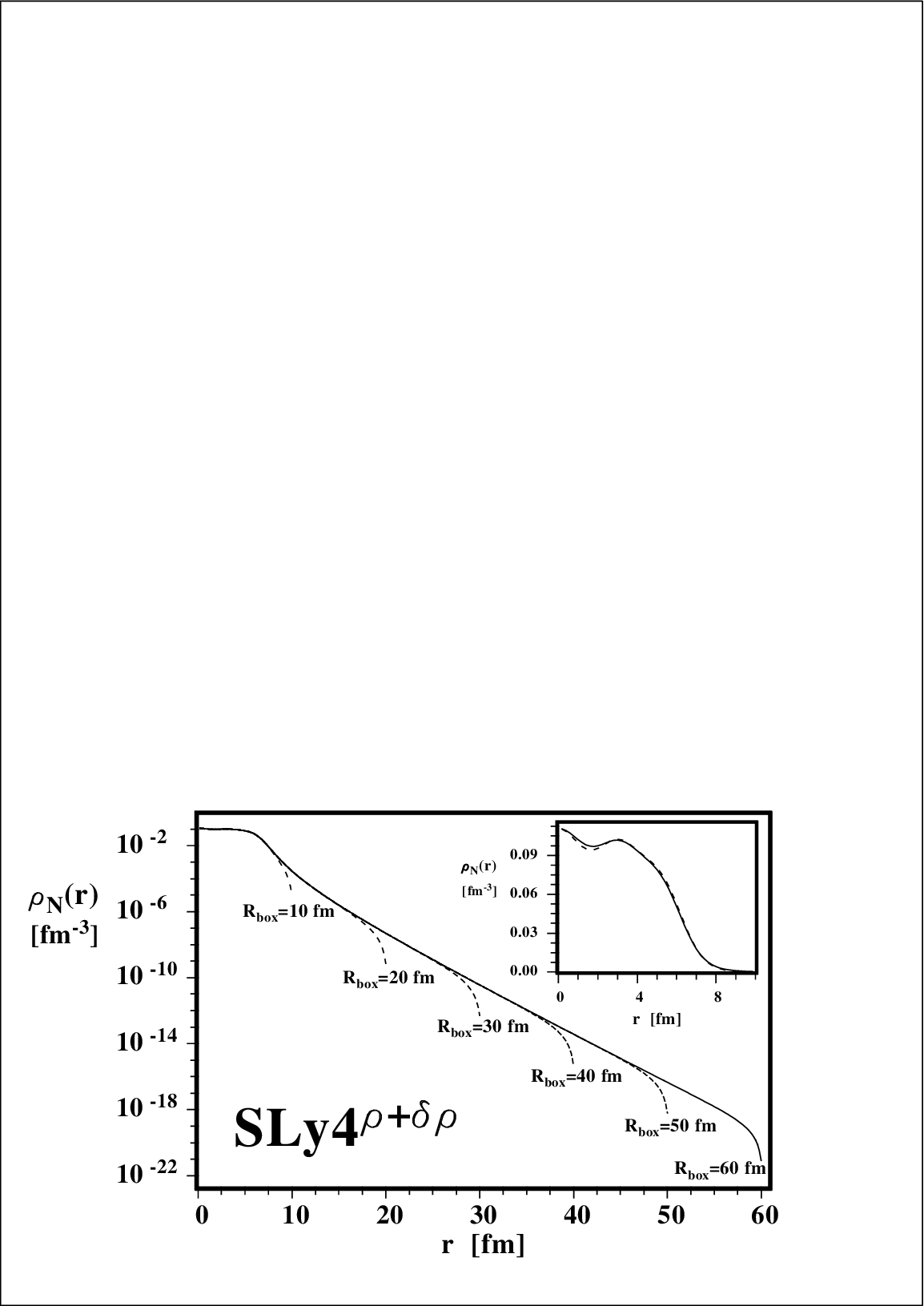} &
\ \includegraphics*[scale=0.42,bb=36 55 487 328]{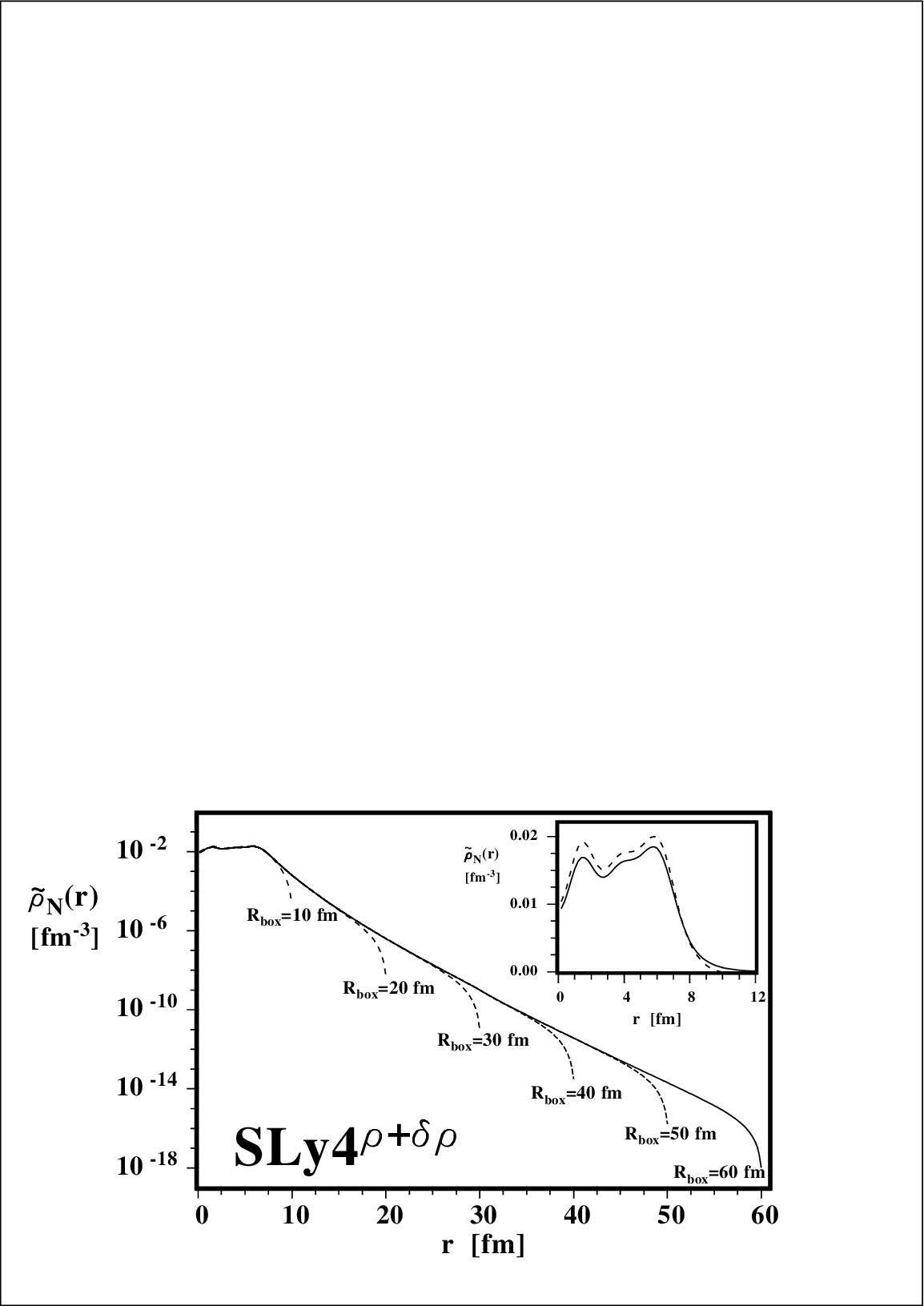} \\
\end{matrix}$
\caption{Neutron particle (left) and pairing (right) densities for different box sizes.
  In the insets the dashed lines correspond to $R_{\mathrm{box}}=10$~fm
  while the solid lines correspond to the other
  values of $R_{\mathrm{box}}$ that cannot be separated in the linear scale.
  \label{fig:f}}
\end{center}
\end{figure}

Neutron particle and pairing densities for different sizes of the box
are shown in Fig.~\ref{fig:f}.
The box radius of 10\,fm is obviously too small and it has only been
included
to show the effect of a very small box on the densities.
For $R_{\mathrm{box}}\geq20$~fm, no differences can be seen in the linear
scale. An interesting fact is that despite the cut that
has been applied for the pairing field at $R_{\mathrm{cut}}=30$~fm,
see Sec.~\ref{sec:instab},
no consequences can be observed on the densities in the asymptotic region.

\subsection{Regularization of the pairing field}
\label{par:exb}

In this section, we present results obtained by using the regularization
method proposed by A.\ Bulgac and Y.\ Yu~\cite{[Bul02]}. Calculations
have been made with the same box size and integration step as
in the previous section, and the partial waves have been included up to $J=43/2$.
The effective force used is SLy4, see table~\ref{tab:paramph},
combined with the mixed pairing force with
parameters given in table~\ref{tab:parampp}.
When evaluating the densities, contributions of quasiparticle states are
included up to the maximum equivalent
energy $E_{\mathrm{max}}$, see Eq.~(\ref{eq:enequiv}),
but once this maximum energy is high enough the
global properties of the nucleus do not depend on it.

\begin{figure}[htbp]
\begin{center}
$\begin{matrix}
\!\mbox{\includegraphics*[scale=0.47,bb=52 60 464 348,clip]{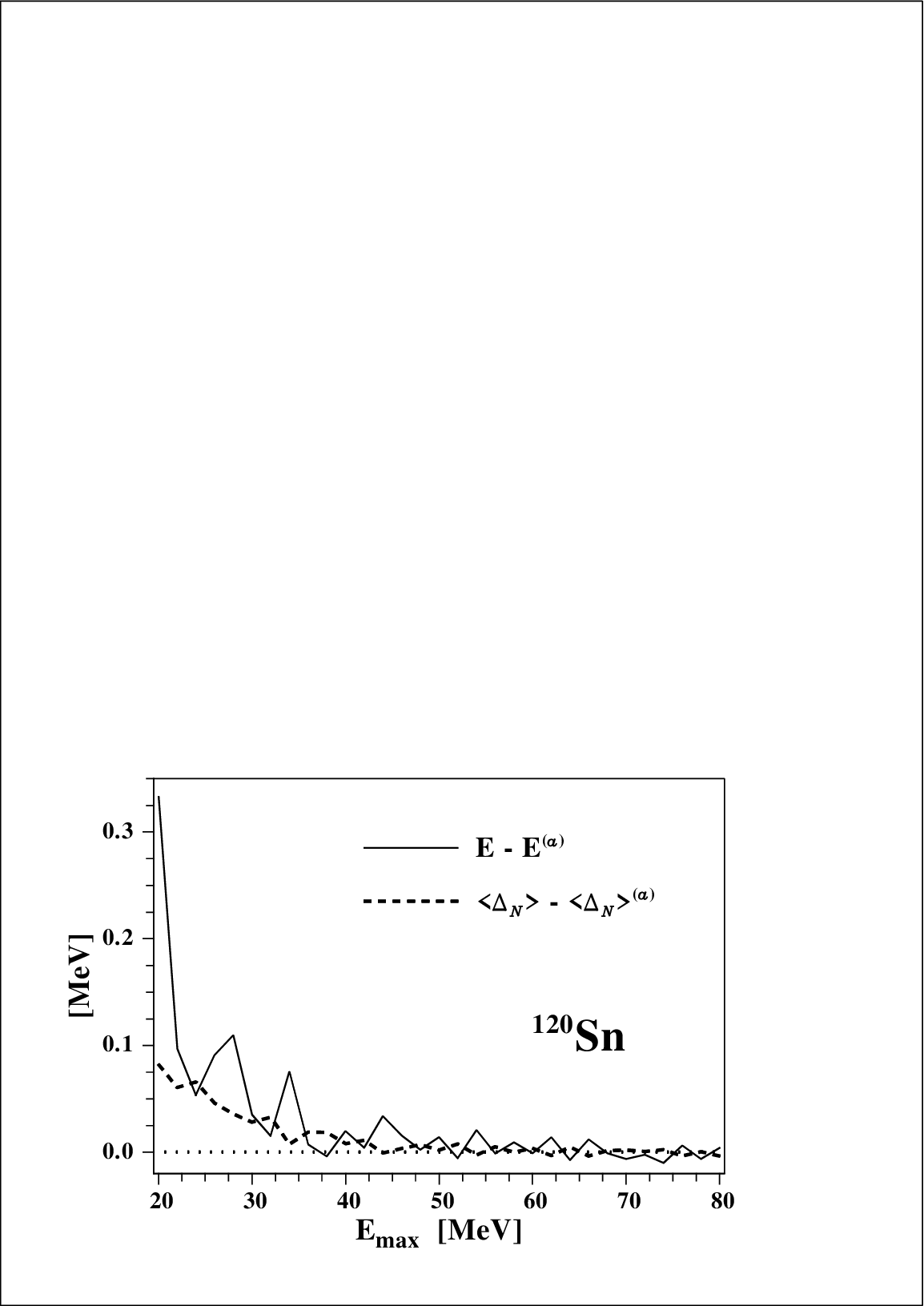}} &
  \mbox{\includegraphics*[scale=0.47,bb=52 60 464 348,clip]{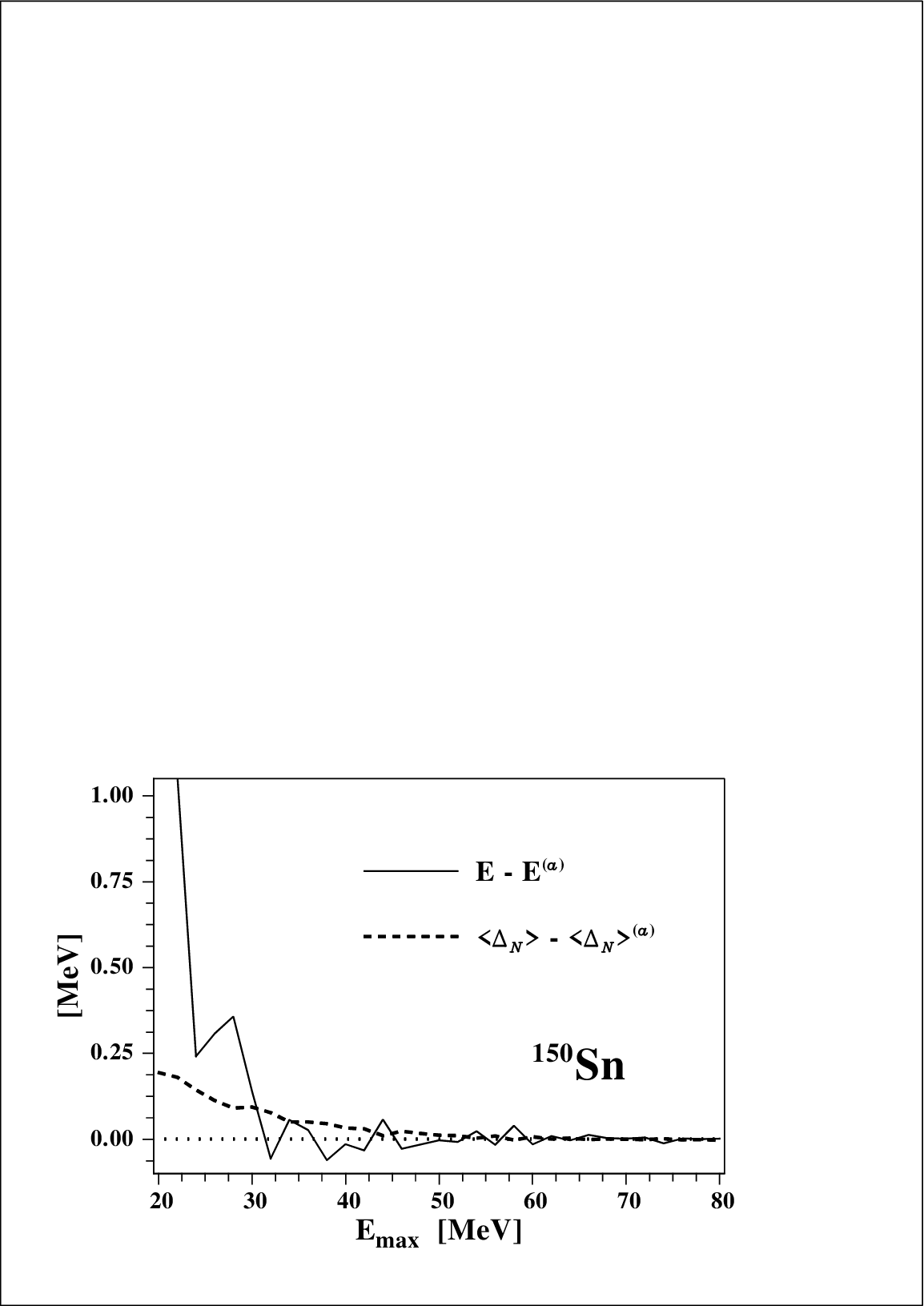}}
\end{matrix}$
\caption{Modulus of the difference between the total energy $E$ and its mean
         value $E^{(a)}$ (solid lines) and between the average neutron
         gap and its mean value (dashed lines) as functions of the
         energy $E_{\mathrm{max}}$ (see text) for $^{120}$Sn (left)
         and $^{150}$Sn (right). The mean values
         are calculated over the interval $60< E_{\mathrm{max}}<80$~MeV.
         \label{fig:cvec}}
\end{center}
\end{figure}

It appears that the stability of results is
very satisfying, even
for a very exotic nucleus. This is shown in Figure~\ref{fig:cvec}, where
the total energy and mean neutron gap are displayed as functions
of $E_{\mathrm{max}}$ for $^{120}$Sn and $^{150}$Sn.
For $E_{\mathrm{max}}$ greater than 60~MeV, where the total energy does
not show any significant evolution, we have evaluated its asymptotic
limit $E^{(a)}$, and the analogous limit of the neutron mean pairing gap
$\langle\Delta\rangle^{(a)}$, by averaging their respective
values over the interval ranging from 60 to 80~MeV. The results are
$E^{(a)}=1\,018.529$~MeV and
$\langle\Delta\rangle^{(a)}=1.245$~MeV for $^{120}$Sn, and
$E^{(a)}=1\,131.492$~MeV and
$\langle\Delta\rangle^{(a)}=1.499$~MeV for $^{150}$Sn.
In this interval, the energies are scattered within $\pm 16$~keV and the
gaps within $\pm 7$~keV.
Since the increase of $E_{\mathrm{max}}$ from 60~MeV to
80~MeV does not change the results in a significant way, the
choice of $E_{\mathrm{max}}=60$~MeV has been made for the rest of this study.
In principle, this value should be readjusted in other mass
region or when using a different effective force.

\begin{figure}[htbp]
\begin{center}
\includegraphics*[scale=0.69,bb=72 62 392 264]{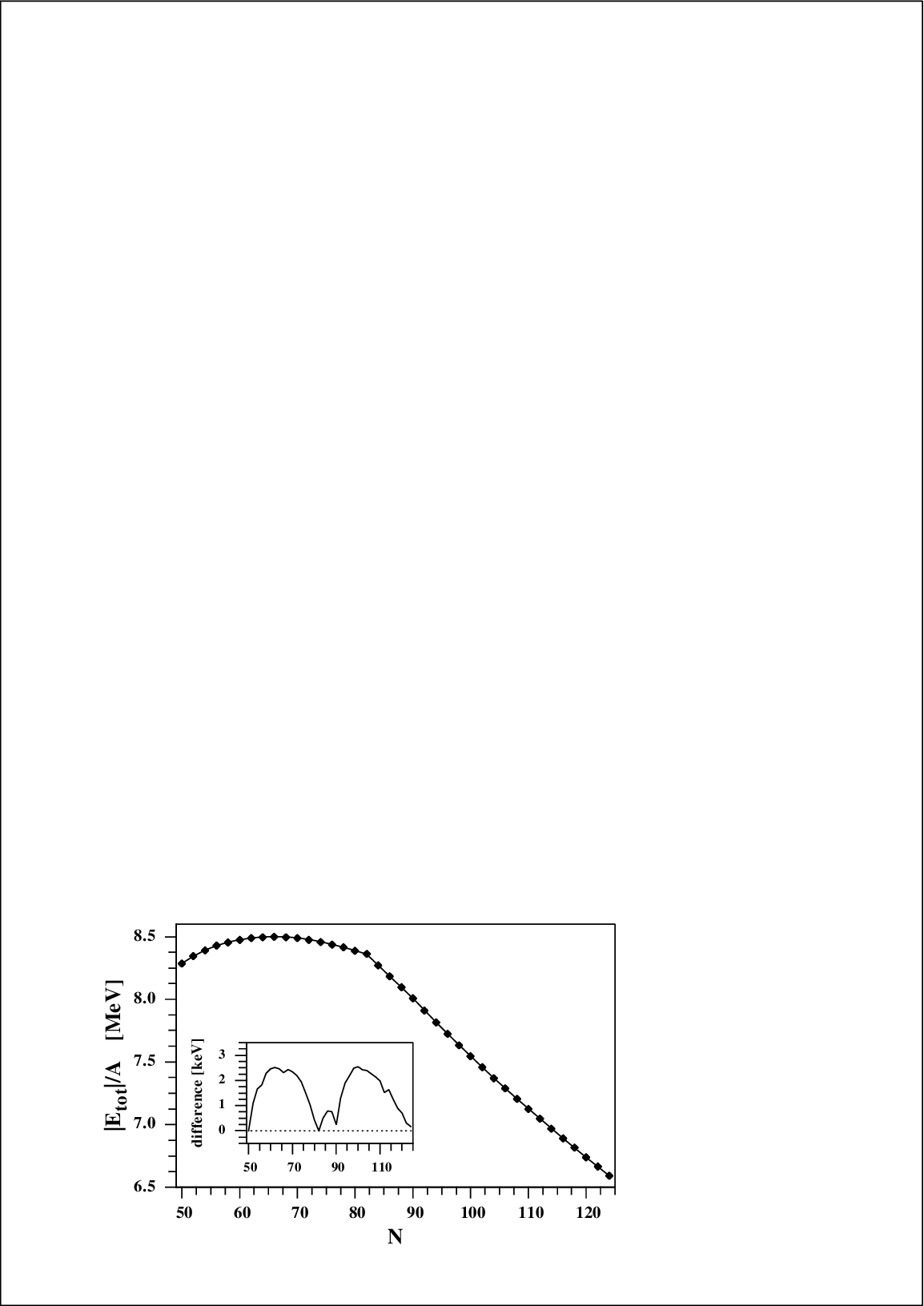}
\includegraphics*[scale=0.69,bb=72 62 392 264]{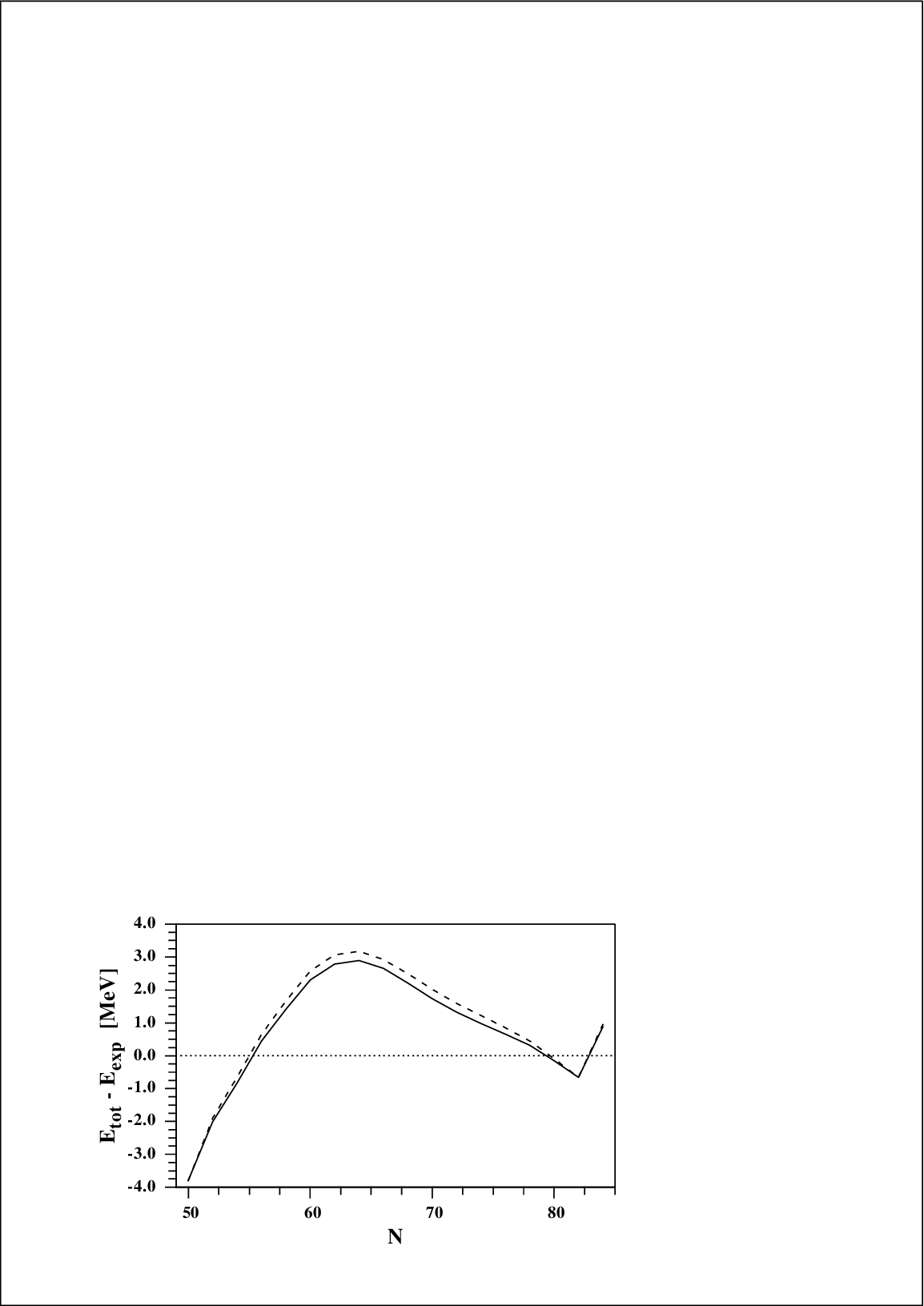}
\caption{Left figure: Binding energy per nucleon in tin isotopes.
The solid line correspond to calculations with a cut-off energy
of $E_{\mathrm{cut}}=60$~MeV in
the equivalent spectrum and the diamond symbols to the regularization
of the energy with the summation of densities up to
$E_{\mathrm{max}}=60$~MeV (see text). The difference between the two sets of
results is plotted in the inset. Right figure: Differences between
the calculated and measured binding energies shown for those tin isotopes
for which the data are known; solid line corresponds
to the introduction of a cut-off and the dashed line to the
regularization.\label{fig:cutvsrege}}
\end{center}
\end{figure}

\begin{figure}[htbp]
\begin{center}
\includegraphics*[scale=0.69,bb=72 62 392 264]{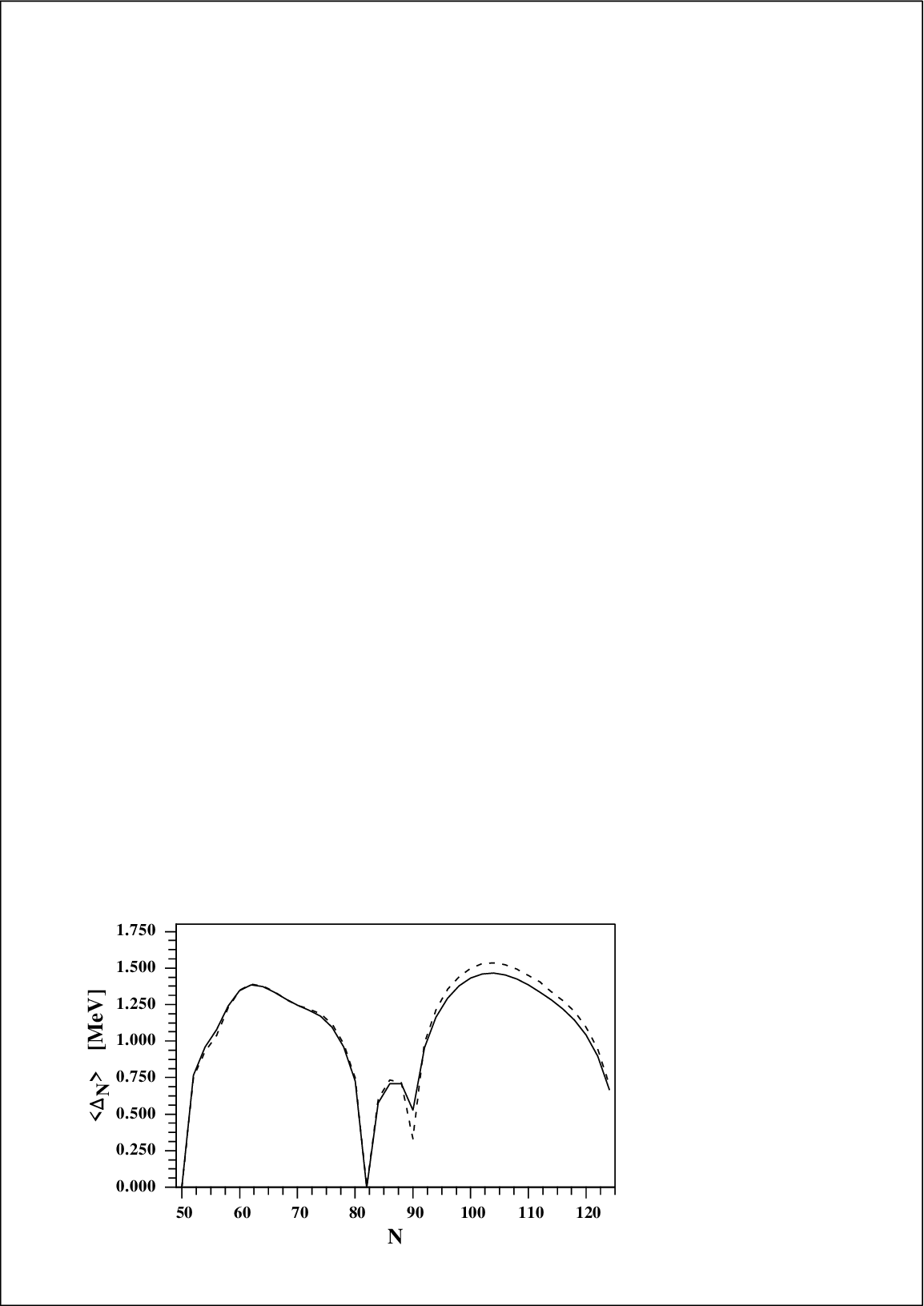}
\includegraphics*[scale=0.69,bb=72 62 392 264]{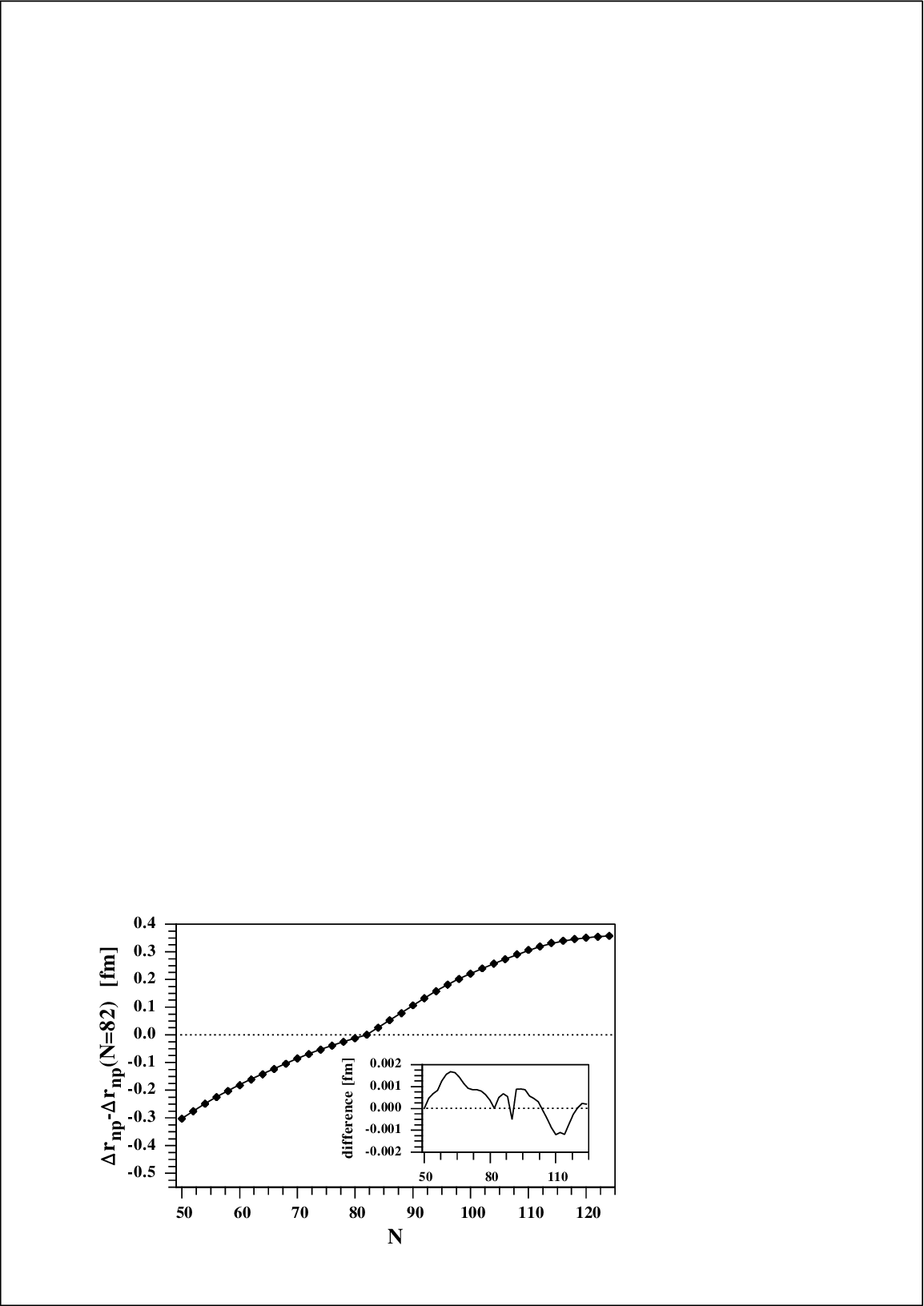}
\caption{Left figure: Mean neutron gap in tin isotopes, the solid line
corresponds to the introduction of a cut-off and the dashed line
to the regularization. Right figure: Differences between the rms radii
of neutrons and protons in tin isotopes compared with the same
quantity for $^{132}$Sn, the solid line corresponds to the introduction
of a cut-off and the diamond symbol to the regularization, while the difference
between the two sets is plotted in the inset.
\label{fig:cutvsregd}}
\end{center}
\end{figure}

Within the cut-off prescription and regularization scheme we have calculated
the series of even-even tin isotopes by using the
SLy4$^{(\rho+\delta\rho)}$ force.
The left part of Fig.~\ref{fig:cutvsrege} displays the
binding energies per particle and the right part the deviation between the calculated binding
energies and the experimental ones~\cite{[Aud03]}. One can see that
both methods give very similar results.
The neutron mean gap are plotted on the left part of
Fig.~\ref{fig:cutvsregd}. In the two sets of calculation,
the strengths of the pairing force have been adjusted
in order to give the same gap in $^{120}$Sn.
Again, we do not observe here
any significant change when using or not the regularization
scheme, although the gap is slightly reduced in heavy
tin isotopes.
Finally, the right part of Fig.~\ref{fig:cutvsregd}
compares the differences between the neutron and proton rms radii
plotted with respect to that in $^{132}$Sn; once again the two methods
give extremely similar results.

\subsection{Neutron drip line}

As a last example we consider the two-neutron drip line for
$Z=48$. The case of $Z=50$
is not interesting for our example because there the neutron pairing correlations vanish
at the two-neutron drip line (at least for the pairing force considered here),
so the residual coupling between the bound and scattering states
disappears.
The two-neutron drip line is defined
by the vanishing Fermi energy, $\lambda_N=0$. In such a system,
the HFB approximation leads to a fully continuous quasiparticle spectrum
and the minimum quasiparticle energy
is determined by the pairing correlations only \cite{[Ben99b]}.
In this extreme situation it is important to check if
the box boundary conditions
have any significant effect on the results. To this end, we have
performed calculations with the box size $R_{\mathrm{box}}$
varying between 10 and 35\,fm and with the regularized pairing
field corresponding to $E_{\mathrm{max}}=60$~MeV.

\begin{figure}[htbp]
\begin{center}
$\begin{matrix}
\!\includegraphics*[scale=0.5,bb=96 53 485 340]{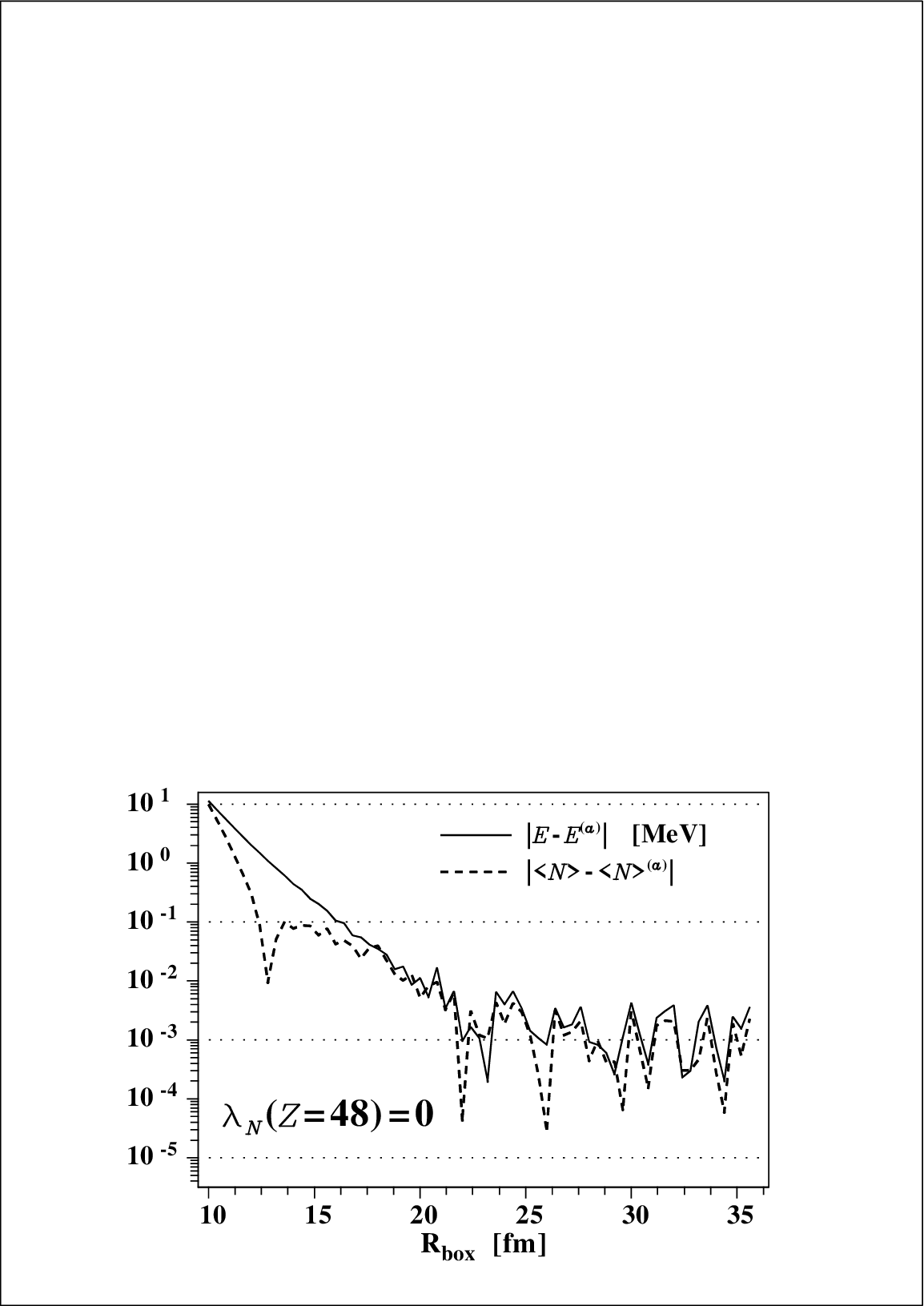} &
  \includegraphics*[scale=0.5,bb=96 53 485 340]{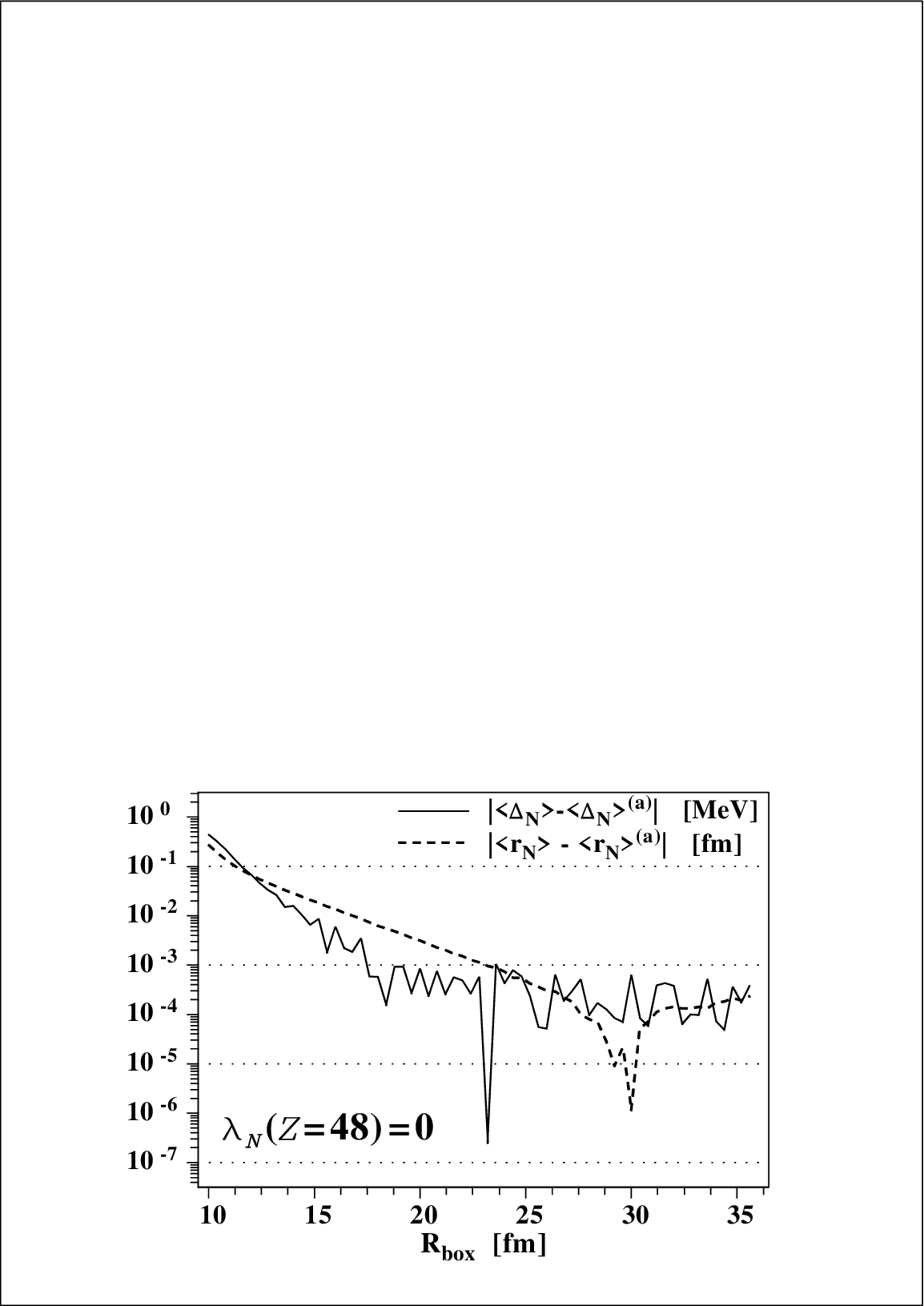} \\
\end{matrix}$
\caption{Differences between the total
  energy, number of neutrons, neutron rms radius, and neutron pairing gap,
  and their corresponding average asymptotic values (see text),
  calculated for
  the $Z=48$ two-neutron drip line nucleus.
         \label{fig:l}}
\end{center}
\end{figure}

We used the same method as in Sec.~\ref{par:exa} to
estimate the average asymptotic values (for $R_{\mathrm{box}}$$\geq$25\,fm)
of the total energy
$E^{(a)}$=$-$1\,095.400\,501\,MeV,
neutron number $\langle n\rangle^{(a)}$=117.053\,163,
pairing gap $\langle \Delta_N\rangle^{(a)}$1.150\,396\,MeV,
and neutron rms radius $\langle r_N\rangle^{(a)}$=5.557\,959\,fm.
Differences of these observables relative to the above average
values are reported in Fig.~\ref{fig:l}.
It is seen that the energy is stable up to several keV once
the radius of the box is bigger than 25\,fm, and no further
significant change is obtained by enlarging it.
The mean neutron number is very
stable too, despite the large spatial extension of the neutron
density in such a drip line nucleus. As discussed in Ref.~\cite{[Ben99b]},
the appearance of a giant neutron halo is prevented
by the pairing correlations (the pairing anti-halo effect) and the neutron rms radius is
perfectly stable with increasing box size. The same is
also true for the neutron pairing gap.

\section{Conclusion}

The code HFBRAD (v\codeversion) which we introduced in this paper
allows for a very rapid determination of self-consistent spherical
ground states of nuclei. It treats the pairing correlations within
the HFB method and implements the regularization of the divergent
part of the pairing energy, and hence is a perfect tool to study
nuclei near drip lines.

Realistic calculations must, of course, take into account the effects
of the deformation, and the codes that are able to perform such tasks
have also been published in Refs. \cite{[Dob04b],[Sto04]}. However,
deformed calculations are generally much more time consuming and
solutions within spherical symmetry will always remain very useful,
not only for test purposes, but also whenever results for many
different values of input parameters are called for. This later aspect
becomes essential when fitting of the force parameters has to be
performed, which will constitute the main future application of
the present code.

\bigskip

\section*{Acknowledgments}

We would like to thank A.~Bulgac and N.~Sandulescu for useful
discussions during the development of the code {HFBRAD}.
This work was supported in part by the Polish Committee for Scientific
Research (KBN) under Contract
No.~1~P03B~059~27 and by the Foundation for Polish Science (FNP);
and by the U.S.\ Department of Energy
under Contract Nos.\ DE-FG02-96ER40963 (University of Tennessee),
DE-AC05-00OR22725 with UT-Battelle, LLC (Oak Ridge National
Laboratory), and DE-FG05-87ER40361 (Joint Institute for Heavy Ion
Research).


\begin{thebibliography}{10}

\bibitem{[Dob04b]}
{J. Dobaczewski and J. Dudek, Comput. Phys. Commun. {\bf 102}, 166 (1997); {\bf
  102}, 183 (1997), {\bf 131}, 164 (2000), J. Dobaczewski and P. Olbratowski,
  {\it ibid.} {\bf 158}, 158 (2004)}.

\bibitem{[Sto04]}
{M.V. Stoitsov, J. Dobaczewski, W. Nazarewicz, and P. Ring, to be published in
  Computer Physics Communication}.

\bibitem{[Vau72]}
{D. Vautherin and D.M. Brink, Phys. Rev. {\bf C5}, 626 (1972)}.

\bibitem{[Bei75]}
{M. Beiner, H. Flocard, N. Van Giai, and P. Quentin, Nucl. Phys. {\bf A238}, 29
  (1975)}.

\bibitem{[Dob84]}
{J. Dobaczewski, H. Flocard and J. Treiner, Nucl. Phys. {\bf A422}, 103
  (1984)}.

\bibitem{[Rei91]}
{P.-G. Reinhard, in {\em Computational Nuclear Physics 1} K. Langanke, J.A.
  Maruhn, and S.E. Koonin, eds. (Springer Verlag, Berlin, 1991) p.28}.

\bibitem{[Pos97a]}
{W. P\"oschl, D. Vretenar, and P. Ring, Comput. Phys. Commun. {\bf 103}, 217
  (1997)}.

\bibitem{[Gra02]}
{M. Grasso, N. Van Giai, and N. Sandulescu, Phys. Lett. B {\bf 535}, 103
  (2002)}.

\bibitem{[Bul02]}
{A. Bulgac and Y. Yu, Phys. Rev. Lett. {\bf 88}, 042504 (2002)}.

\bibitem{[RS80]}
{P. Ring and P. Schuck, {\sl The Nuclear Many-Body Problem} (Springer-Verlag,
  Berlin, 1980)}.

\bibitem{[Eng75]}
{Y.M. Engel, D.M. Brink, K. Goeke, S.J. Krieger, and D. Vautherin, Nucl. Phys.
  {\bf A249}, 215 (1975)}.

\bibitem{[Ben00]}
{K. Bennaceur, J. Dobaczewski, and M. P{\l}oszajczak, Phys. Lett. {\bf B496},
  154 (2000)}.

\bibitem{[Dob95]}
{J. Dobaczewski and J. Dudek, Phys. Rev. {\bf C52}, 1827 (1995)}.

\bibitem{[Neg72]}
{J.W. Negele and D. Vautherin, Phys. Rev. {\bf C5}, 1472 (1972)}.

\bibitem{[Tit74]}
{C. Titin-Schnaider and P. Quentin, Phys. Lett. {\bf 49B}, 397 (1974)}.

\bibitem{[Ska01]}
{J. Skalski, Phys. Rev. {\bf C63}, 024312 (2001)}.

\bibitem{[Bul80]}
{A. Bulgac, Preprint FT-194-1980, Central Institute of Physics, Bucharest,
  1980; nucl-th/9907088}.

\bibitem{[Dob96]}
{J. Dobaczewski, W. Nazarewicz, T.R. Werner, J.-F. Berger, C.R. Chinn, and J.
  Decharg\'e, Phys. Rev. {\bf C53}, 2809 (1996)}.

\bibitem{[Bul01c]}
{A. Bulgac and Y. Yu, Beijing International Summer School on Subatomic Physics,
  2001 and International Workshop on Nuclear Physics with RIB, Lanzhou, 2001,
  nucl-th/0109083}.

\bibitem{[Ben03]}
{M. Bender, P.-H. Heenen, and P.-G. Reinhard, Rev. Mod. Phys. {\bf 75}, 121
  (2003)}.

\bibitem{[Bar82]}
{J. Bartel, P. Quentin, M. Brack, C. Guet, and H.B. H{\aa}kansson, Nucl. Phys.
  {\bf A386}, 79 (1982)}.

\bibitem{[Cha98]}
{E. Chabanat, P. Bonche, P. Haensel, J. Meyer, and F. Schaeffer, Nucl. Phys.
  {\bf A635}, 231 (1998)}.

\bibitem{[Dah74]}
{G. Dahlquist and \AA. Bj\"orck, Numerical Methods, (Prentice-Hall, Engelwoods
  Cliffs, 1974)}.

\bibitem{[Bel87]}
{S.T. Belyaev, A.V. Smirnov, S.V. Tolokonnikov, and S.A. Fayans, Sov. J. Nucl.
  Phys. {\bf 45}, 783 (1987)}.

\bibitem{[Aud03]}
{G. Audi, A.H. Wapstra, and C. Thibault, Nucl. Phys. {\bf A729}, 337 (2003)}.

\bibitem{[Ben99b]}
{K. Bennaceur, J. Dobaczewski, and M. P{\l}oszajczak, Phys. Rev. {\bf C60},
  034308 (1999)}.

\end{thebibliography}

\clearpage

{\bf\large TEST RUN INPUT}
{\tt\footnotesize
\baselineskip 1ex
\begin{scriptsize}
\begin{verbatim}
&input
force = "SLY4",         mesh_points = 150,
integ_step = 0.2,       it_max = 150,
eps_energy = 1.e-9,     max_delta = 1.e-7,
boundary_condition = 0, xmu = 0.65,
bogolyubov = T, T,      pairing_force = 3,   regularization = F /

&nucleus  neutron = 100, proton = 50, j_max = 39, 25 /

\end{verbatim}
\end{scriptsize}
}

\clearpage

{\bf\large TEST RUN OUTPUT}

{\tt\scriptsize
\baselineskip 1ex

\begin{verbatim}

********************************** PR =  8 ***********************************


                            1 nucleus to compute

                             Skyrme Force = SLY4

                      --  N = 100    --    Z = 50    --

 --------------------------------  INPUT DATA  ---------------------------------
  it_max = 150       eps energ. = 1.000E-09       max delta = 1.000E-07
  Boundary_condition = 0
  Pairing force: Volume + Surface pairing  Bogolyubov transf. for both isospins
  npt = 150     integ. step = 0.200 fm     Rbox = 30.00 fm     r_cut =  30.00 fm
 -------------------------------------------------------------------------------
 Skyrme force parameters:
      t0 = -2488.913000   x0 =   0.834000    t1 =   486.818000   x1 =  -0.344000
      t2 =  -546.395000   x2 =  -1.000000    t3 = 13777.000000   x3 =   1.354000
   gamma =     0.166667    W =   123.000000
     t0' =  -283.330000  t3' =  5312.437500   gamma' =     1.000000
  hb2/2m = 20.7355300   (with 1-body c.m. correction)
 -------------------------------------------------------------------------------
  xmu = 0.65       2Jmax = 39 (neutrons),   25 (protons)
  cut_off =  60.000 MeV  (with diff.  1.000 MeV)      energ. step =  0.120 MeV
 -------------------------------------------------------------------------------

 iter  Nint       Fermi Energies           Mean Delta       Total Energy  | D(E)/E |
 ---- ------ ---- N --------- Z ---- ---- N ------- Z ---- -------------- ----------
   1   46225 -10.134865  -26.280381   0.512035   0.119232  -1066.9271332  0.1000E+01
   2   50871  -6.319108  -23.574948   0.669110   0.107073  -1107.3301783  0.3649E-01
   3   48787  -4.137505  -21.419928   0.808457   0.094809  -1155.7802126  0.4192E-01
   4   47484  -2.618452  -20.114633   0.930301   0.083488  -1127.5652230  0.2502E-01
   5   42335  -1.815561  -19.298385   1.024427   0.073335  -1129.0685156  0.1331E-02
   6   41349  -1.442125  -18.713186   1.095235   0.064446  -1137.4968802  0.7410E-02
   7   27489  -1.151068  -18.353445   1.150466   0.056740  -1132.5657761  0.4354E-02
   8   26609  -1.005877  -18.095592   1.194775   0.050069  -1131.4434353  0.9920E-03
   9   28014  -0.935437  -17.921837   1.230628   0.044285  -1131.4978982  0.4813E-04
  10   27079  -0.907358  -17.807723   1.259893   0.039286  -1131.5828931  0.7511E-04
  11   24808  -0.901746  -17.734617   1.284171   0.034992  -1131.6252463  0.3743E-04
  12   21131  -0.908367  -17.688935   1.304579   0.031240  -1131.6516680  0.2335E-04
  13   20002  -0.921223  -17.661557   1.321914   0.027931  -1131.6794938  0.2459E-04
  14   19279  -0.936657  -17.646425   1.336752   0.025005  -1131.7082567  0.2542E-04
  15   19266  -0.952516  -17.639172   1.349523   0.022408  -1131.7340677  0.2281E-04
  16   19106  -0.967636  -17.636997   1.360551   0.020098  -1131.7555177  0.1895E-04
  17   19132  -0.981457  -17.637944   1.370095   0.018038  -1131.7729552  0.1541E-04
  18   17388  -0.993767  -17.640713   1.378361   0.016199  -1131.7871750  0.1256E-04
  19   17927  -1.004548  -17.644451   1.385525   0.014554  -1131.7988684  0.1033E-04
  20   18494  -1.013882  -17.648609   1.391732   0.013081  -1131.8086476  0.8640E-05
  21   18789  -1.021901  -17.652842   1.397107   0.011760  -1131.8168763  0.7270E-05
  22   18723  -1.028753  -17.656939   1.401761   0.010576  -1131.8238454  0.6157E-05
  23   18735  -1.034587  -17.660780   1.405788   0.009512  -1131.8297696  0.5234E-05
  24   18743  -1.039540  -17.664305   1.409269   0.008557  -1131.8348142  0.4457E-05
  25   18669  -1.043739  -17.667489   1.412277   0.007699  -1131.8391107  0.3796E-05
  26   18629  -1.047294  -17.670334   1.414875   0.006927  -1131.8427681  0.3231E-05
  27   18509  -1.050303  -17.672854   1.417117   0.006234  -1131.8458785  0.2748E-05
  28   18488  -1.052848  -17.675071   1.419051   0.005610  -1131.8485207  0.2334E-05
  29   18497  -1.055000  -17.677011   1.420718   0.005049  -1131.8507625  0.1981E-05
  30   18454  -1.056821  -17.678703   1.422155   0.004544  -1131.8526628  0.1679E-05
  31   18286  -1.058362  -17.680172   1.423393   0.004090  -1131.8542722  0.1422E-05
  32   17843  -1.059666  -17.681446   1.424459   0.003682  -1131.8556344  0.1204E-05
  33   16086  -1.060769  -17.682547   1.425376   0.003314  -1131.8567870  0.1018E-05
  34   11224  -1.061704  -17.683497   1.426166   0.002983  -1131.8577619  0.8613E-06
  35   11080  -1.062497  -17.684315   1.426845   0.002685  -1131.8585863  0.7284E-06
  36   11117  -1.063168  -17.685020   1.427429   0.002417  -1131.8592836  0.6160E-06
  37   11154  -1.063737  -17.685625   1.427932   0.002176  -1131.8598733  0.5210E-06
  38   11036  -1.064220  -17.686145   1.428363   0.001958  -1131.8603721  0.4407E-06
  39   11135  -1.064630  -17.686591   1.428734   0.001763  -1131.8607941  0.3728E-06
  40   11143  -1.064978  -17.686973   1.429053   0.001587  -1131.8611513  0.3156E-06
  41   11136  -1.065273  -17.687301   1.429327   0.001429  -1131.8614538  0.2673E-06
  42   11188  -1.065524  -17.687582   1.429563   0.001286  -1131.8617099  0.2263E-06
  43   11240  -1.065738  -17.687823   1.429765   0.001158  -1131.8619269  0.1917E-06
  44   11129  -1.065919  -17.688029   1.429938   0.001042  -1131.8621109  0.1625E-06
  45   11274  -1.066073  -17.688205   1.430087   0.000938  -1131.8622668  0.1378E-06
  46   11204  -1.066204  -17.688355   1.430215   0.000845  -1131.8623992  0.1170E-06
  47   11186  -1.066316  -17.688484   1.430325   0.000760  -1131.8625113  0.9910E-07
  48   11235  -1.066411  -17.688594   1.430419   0.000684  -1131.8626065  0.8409E-07
  49   11199  -1.066492  -17.688688   1.430500   0.000616  -1131.8626873  0.7137E-07
  50   11214  -1.066560  -17.688769   1.430570   0.000555  -1131.8627560  0.6072E-07
  51   11175  -1.066619  -17.688838   1.430629   0.000499  -1131.8628142  0.5140E-07
  52   11129  -1.066669  -17.688897   1.430680   0.000449  -1131.8628638  0.4381E-07
  53   11240  -1.066711  -17.688946   1.430724   0.000405  -1131.8629060  0.3726E-07
  54   11181  -1.066747  -17.688990   1.430762   0.000364  -1131.8629417  0.3158E-07
  55   11248  -1.066778  -17.689027   1.430794   0.000328  -1131.8629722  0.2695E-07
  56   11161  -1.066805  -17.689058   1.430822   0.000295  -1131.8629981  0.2287E-07
  57   11200  -1.066827  -17.689084   1.430846   0.000266  -1131.8630201  0.1940E-07
  58   11209  -1.066846  -17.689107   1.430866   0.000239  -1131.8630389  0.1661E-07
  59   11213  -1.066862  -17.689128   1.430883   0.000215  -1131.8630548  0.1407E-07
  60   11203  -1.066876  -17.689145   1.430898   0.000194  -1131.8630684  0.1202E-07
  61   11201  -1.066888  -17.689157   1.430911   0.000174  -1131.8630799  0.1020E-07
  62   11241  -1.066898  -17.689170   1.430922   0.000157  -1131.8630898  0.8715E-08
  63   11216  -1.066907  -17.689180   1.430932   0.000141  -1131.8630982  0.7434E-08
  64   11236  -1.066914  -17.689186   1.430940   0.000127  -1131.8631053  0.6255E-08
  65   11250  -1.066920  -17.689190   1.430947   0.000115  -1131.8631114  0.5421E-08
  66   11299  -1.066926  -17.689203   1.430953   0.000103  -1131.8631166  0.4555E-08
  67   11210  -1.066930  -17.689209   1.430958   0.000093  -1131.8631210  0.3932E-08
  68   11196  -1.066934  -17.689211   1.430962   0.000084  -1131.8631248  0.3321E-08
  69   11180  -1.066937  -17.689203   1.430966   0.000075  -1131.8631280  0.2858E-08
  70   11202  -1.066940  -17.689218   1.430969   0.000068  -1131.8631308  0.2430E-08
  71   11207  -1.066943  -17.689218   1.430972   0.000061  -1131.8631331  0.2067E-08
  72   11229  -1.066945  -17.689231   1.430974   0.000055  -1131.8631351  0.1734E-08
  73   11192  -1.066946  -17.689211   1.430976   0.000049  -1131.8631368  0.1492E-08
  74   11258  -1.066948  -17.689249   1.430978   0.000044  -1131.8631382  0.1261E-08
  75   11192  -1.066949  -17.689249   1.430979   0.000040  -1131.8631395  0.1116E-08
   warning: Iterations are not progressing well and pairing seems to be small...
            Trying to force convergence for it = 2
  76   11211  -1.066950  -17.652243   1.430981   0.000002  -1131.8631406  0.9561E-09
  77   14395  -1.066951  -17.648398   1.430982   0.000001  -1131.8631415  0.7948E-09
  78   14316  -1.066952  -17.653151   1.430983   0.000001  -1131.8631421  0.5991E-09
  79   14417  -1.066953  -17.653151   1.430983   0.000001  -1131.8631428  0.6288E-09
  80   11225  -1.066953  -17.631479   1.430984   0.000001  -1131.8631434  0.4951E-09
  81   14323  -1.066954  -17.640449   1.430985   0.000001  -1131.8631438  0.3779E-09
  82   14242  -1.066954  -17.640449   1.430985   0.000001  -1131.8631442  0.3551E-09
  83   11168  -1.066955  -17.640449   1.430986   0.000001  -1131.8631445  0.2583E-09
  84   11225  -1.066955  -17.623581   1.430986   0.000001  -1131.8631449  0.3381E-09
  85   14356  -1.066955  -17.581957   1.430986   0.000001  -1131.8631452  0.2942E-09
  86   14393  -1.066955  -17.581957   1.430987   0.000001  -1131.8631454  0.1300E-09
  87   11269  -1.066956  -17.645585   1.430987   0.000001  -1131.8631455  0.1205E-09
  88   14414  -1.066956  -17.606532   1.430987   0.000000  -1131.8631457  0.1428E-09
  89   14353  -1.066956  -17.606532   1.430987   0.000000  -1131.8631459  0.1478E-09

 iter  Nint       Fermi Energies           Mean Delta       Total Energy  | D(E)/E |
 ---- ------ ---- N --------- Z ---- ---- N ------- Z ---- -------------- ----------
  90   11162  -1.066956  -17.486178   1.430987   0.000000  -1131.8631460  0.1546E-09

                     Neutrons        Protons          Total
 ..............................................................
 Fermi Ener.  =     -1.06695605    -17.48617778
 Mean Gaps    =     -1.43098737     -0.00000038
 Kinetic En.  =   1987.30592234    758.56803134   2745.87395368
 Pairing En.  =    -22.61435165      0.00000000    -22.61435165
 Pair.Kin.En. =      0.00000000      0.00000000      0.00000000

 Energies (in MeV):
  E/A =  -7.545754    Variat. =   0.000000      =====>  Etot = -1131.863146
 Contributions:
  Field = -4118.250650  Spin-Or. =  -67.645555     Coul. =  349.004642
 (Rear. =   777.050463)                         Coul.Ex. =  -18.231185

                    Neutrons       Protons          Total         Charge
 Part. numbers:   100.000000      50.000000
 Radii:             5.263562       4.820502       5.120137       4.886434

  ----------------------------------------------------------------------------------

\end{verbatim}
}

\clearpage

\end{document}